\documentclass[dvips, 12pt]{article}

\parskip \medskipamount

\usepackage{amsmath, amssymb, amsthm, makeidx, graphicx, esint}

\title{Hierarchical organization versus self-organization}

\author{Evo Busseniers \\
Global Brain Institute\\
Vrije Universiteit Brussel}

\begin{document}
\maketitle  


\section{Introduction}

In this paper we try to define the difference between hierarchical organization and self-organization. Organization is defined as a structure with a function. So we can define the difference between hierarchical organization and self-organization both on the structure as on the function. In the next two chapters these two definitions are given. For the structure we will use some existing definitions in graph theory, for the function we will use existing theory on (self-)organization. In the third chapter we will look how these two definitions agree. Finally we give a conclusion.

\section{Structure}
We will only look at structures who can be represented as a graph. A graph $G$ is defined as the set of two sets $V$ and $E$. $V$ is the set of all nodes, and $E$ is the set of all edges. We can represent an edge as a pair of nodes, since an edge connects two nodes. In this paper we will only work with undirected graphs. In a directed graph there is a starting node and an ending node of an edge, so an edge is represented as a couple of two nodes. We can define some properties of a node, which depends on the graph structure. By these properties we can define hierarchy. 

  \subsection{Degree}
\begin{figure}[htbf]
\begin{center}
\includegraphics[scale=0.3]{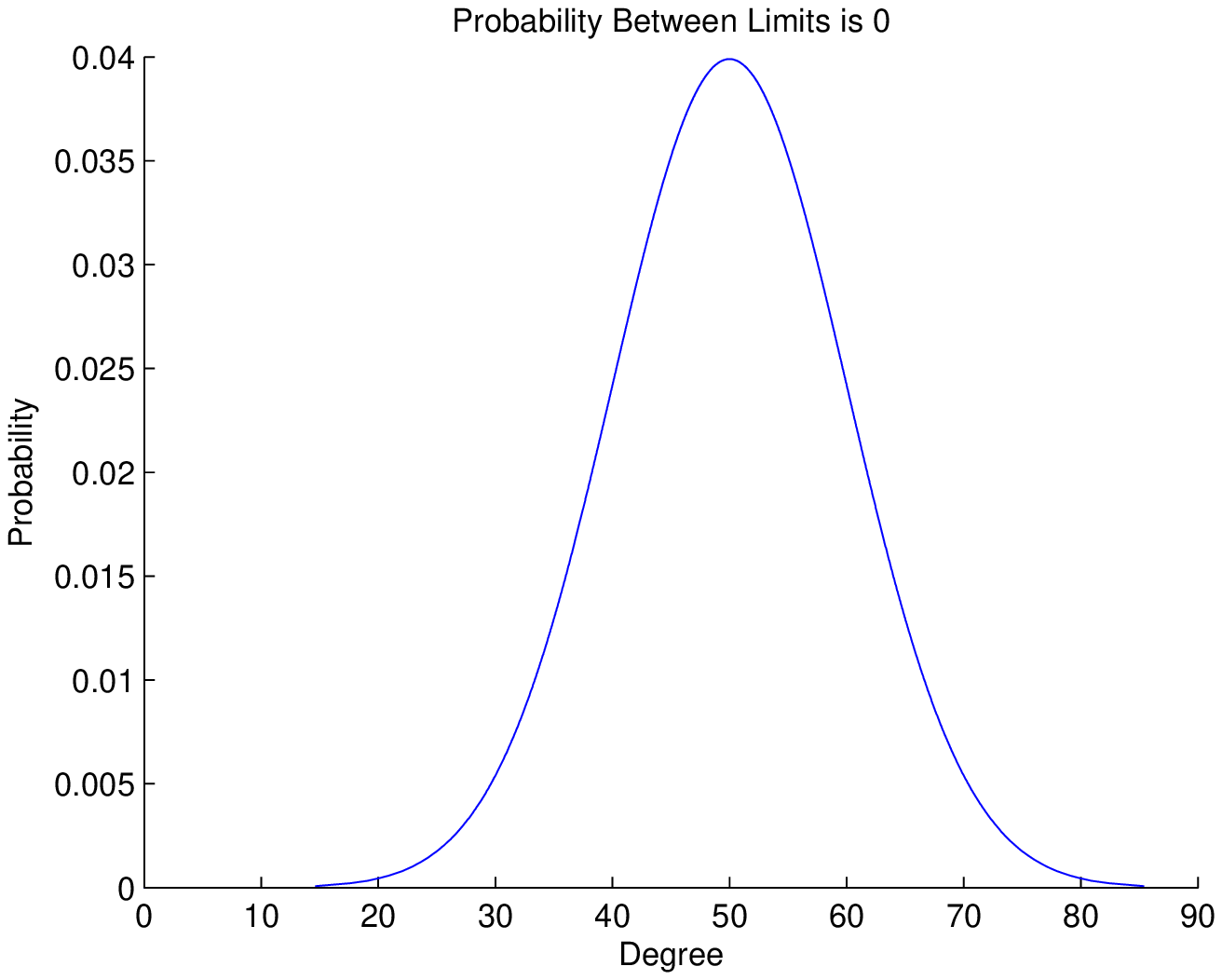}
\includegraphics[scale=0.3]{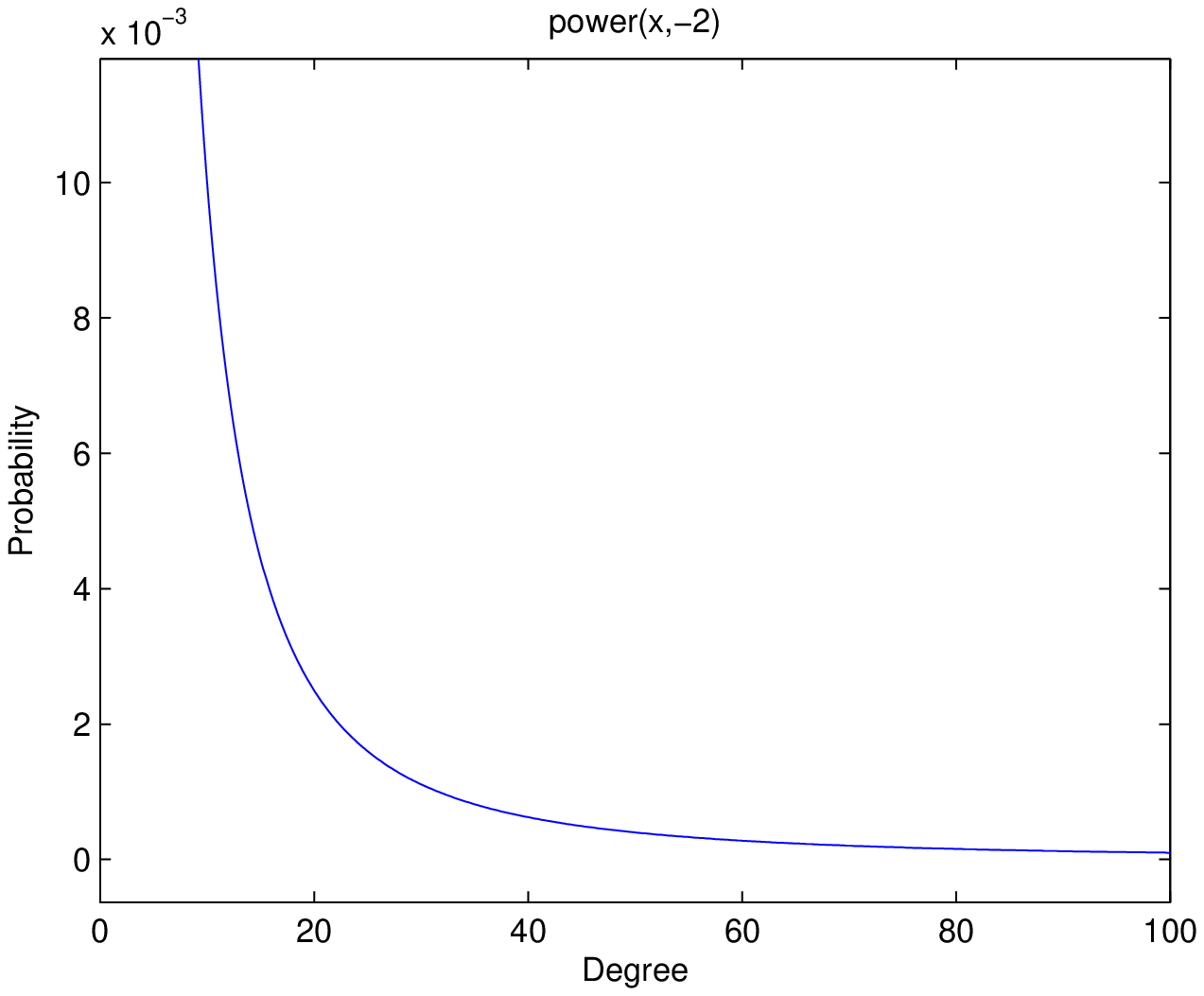}\\
\hspace{4.6cm}\includegraphics[scale=0.3]{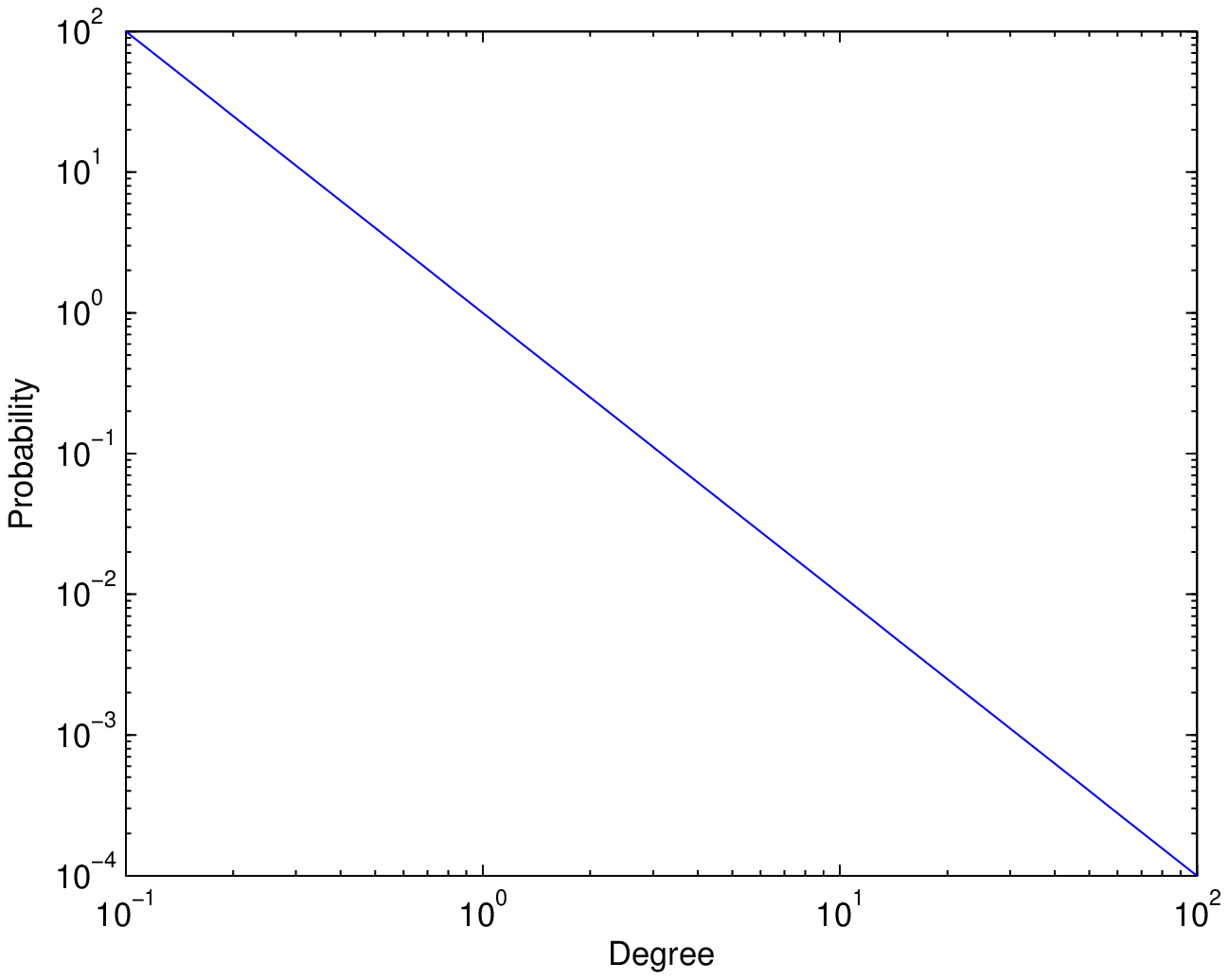}
\caption{Probability distribution of random and scale-free network, second line on loglog-scale}\label{prob}
\end{center}
\end{figure}

We define the degree of a node $v$ as the number of edges that contain $v$. This is the same as the number of neighbors of $v$, where a neighbor is a node which is connected by an edge with $v$. We can now look at how the degree is distributed over the network: whether most nodes have the same degree or some nodes have a higher degree then others. We do that by looking at the probability distribution $P$, where $P(k)$ is the probability that a node has degree $k$. Through this distribution, two sorts of networks are defined: a random network and a scale-free network. In a random network $P$ follows a normal distribution: nodes vary around the mean, the further away from the mean, the less nodes with this degree. In a scale-free network  $P$ follows a power-law:  
	\[P(k)\sim k^{-\lambda} ,
\]
with $\lambda$ some constant\cite{bollobas_random_2001, barabasi_scale-free_2003, newman_structure_2003}. So most nodes have a low degree, there are only a few nodes with a high degree. See figure \ref{prob} for the plots of the distributions. 
A rather intuitive way to define hierarchy, would be to look at the distribution of the degree. The idea is that nodes with a higher degree, have more influence. By that assumption, a hierarchical network could be defined as a scale-free network, and a non-hierarchical network as a random network. But this assumption is wrong, since a node could have only a few neighbors, but still have a lot of influence since it connects different clusters. For this reason, we will work with a better property: the cluster coefficient. 
 \subsection{Cluster coefficient}
The cluster coefficient is a measurement for how good the neighbors of a certain node are connected. The idea behind using it to define hierarchy, is that if the neighbors of a certain node are not that good connected, they depend more on that node. It will happen more that communication has to pass via that certain node to reach a neighbor. If the cluster coefficient of a node is high, it is interchangeable with its neighbors, since the connections of a neighbor are pretty similar with the connections of our starting node.   The cluster coefficient $C(v)$ of a node $v$ is defined as
\begin{eqnarray}
C(v)&=& \frac{|\mbox{edges between neighbors}|}{|\mbox{total possible edges between neighbors}|} \nonumber \\
    &=& \frac{n_v}{\frac{k(k-1)}{2}} \nonumber
\end{eqnarray}
with $n_v=|\mbox{edges between neighbors}|$; $k=$number of neighbors. With this measurement, we can define two different networks\cite{bollobas_random_2001}. The definitions are a refinement of the previous definition of scale-free network. A non-hierarchical network is a network where the cluster coefficient is independent of the degree; the averages of the cluster coefficients of all nodes with the same degree, is (approximately) the same for all the degrees. In a hierarchical network, the higher the degree of a node is, the lower the cluster coefficient. Here the cluster coefficient follows the scaling law

	\[C(k) \sim k^{-1}
\]
with $k$ the degree, and $C(k)$ the average cluster coefficient of all nodes with degree $k$ \cite{barabasi_scale-free_2003}. 

 \subsection{The three networks}
The following figure gives a good summary about the differences the three networks have:

\begin{figure}[htbf]
\begin{center}
\includegraphics[scale=0.3]{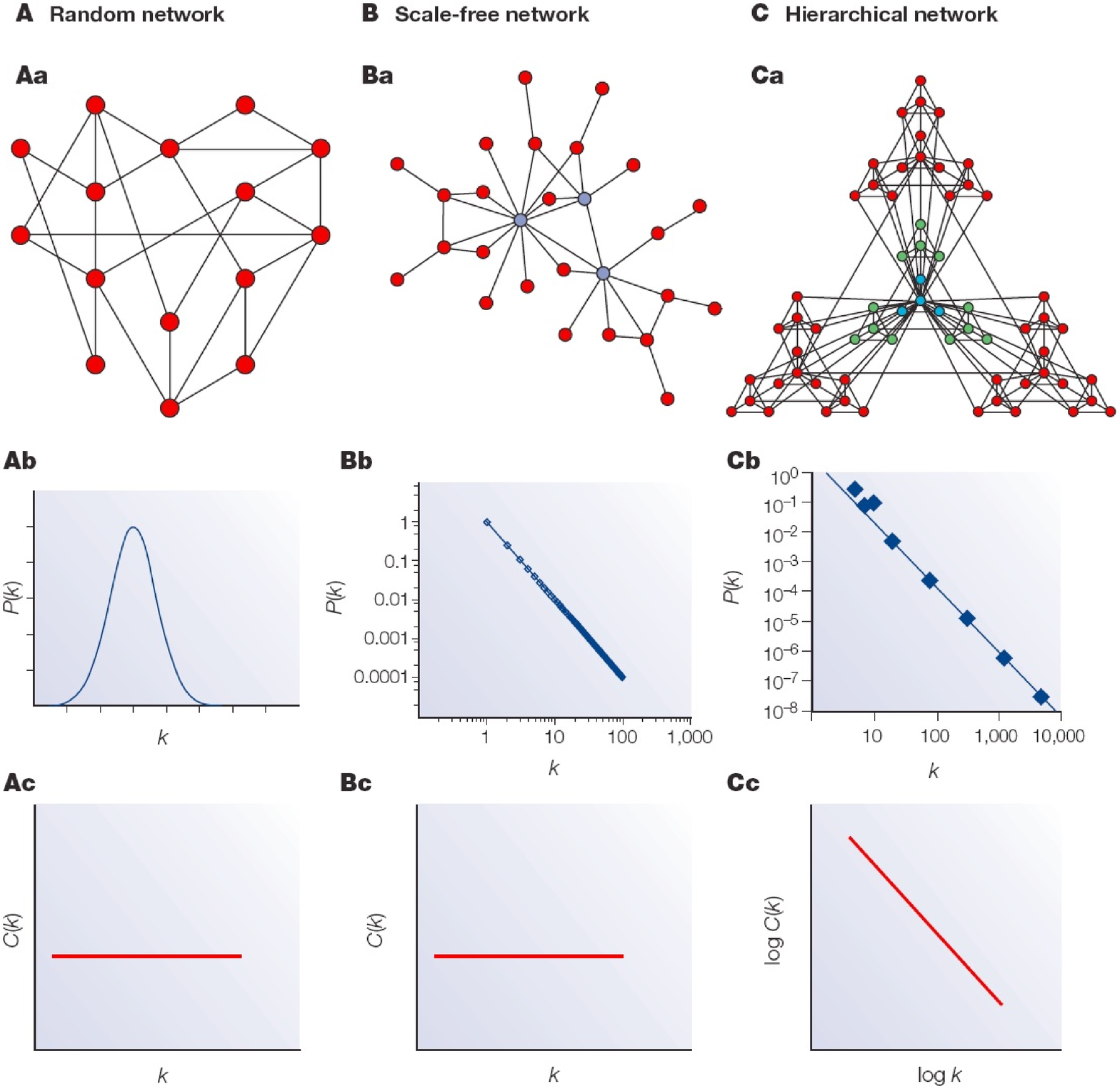}
\caption{Probability distribution and cluster coefficient against degree in 3 networks, from \cite{barabasi_network_2004}}
\end{center}
\end{figure}

We now try to construct examples of these networks, with an equal number of nodes and an approximately equal number of edges. A random network is easily constructed:   you start by the number of nodes wanted, unconnected. Then, two nodes are selected randomly and connected by an edge. This step is repeated until the graph has the number of edges that was desired. \\
The hierarchical network used here, was based on this figure:\\
\begin{figure}[htbf]
\begin{center}
\includegraphics[scale=0.5]{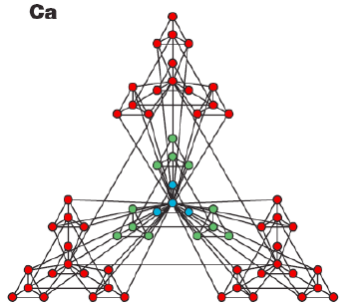}
\caption{Hierarchical network, from \cite{barabasi_network_2004}}
\end{center}
\end{figure}

Formally, clusters are used as building stones. At every scale, a bigger cluster is constructed out of the existing clusters, the building stones. One of these clusters is selected as being the leading cluster, its leader is also the leader of the bigger cluster. This leader is connected with everything in the bigger cluster. The leaders of the smaller clusters (the building blocks) are also connected. Here we use a scaling degree of 3, by which we mean that each leader is connected with 3 other clusters, thus each cluster consists of 4 building blocks. We thus start with 4 clusters, existing of one node. These four nodes form a cluster on a higher scale. One of these nodes is the leading node of the cluster, which is connected to every node, the other nodes are also connected with each other, as they are the leaders of clusters (they are leaders of themselves) being together in a cluster of a higher scale. So in this scale, we just have a complete graph with 4 nodes. We now construct 4 of these clusters, which will become a cluster of a higher scale. One of these clusters will be the leading cluster, with its leading node the leader of the big cluster. This node will be connected with everything. The leaders of the other clusters will be connected with each other. This procedure is repeated for a number of steps, depending of the scale the hierarchical network has to have. In this paper we will use a network of $121$ nodes and $1025$ edges. For this reason we also build our random network with $121$ nodes and $1025$ edges.\\
To create a non-hierarchical network, we use the B-A algorithm \cite[p. 71]{albert_statistical_2001}. This works as follows: we start from a small graph, nodes are added with $m$ edges ($m$ is a parameter of the system, mostly between 1 and 5). But these connections don't happen randomly: there is more chance to connect with nodes who have already a high degree. This is called preferential attachment. Formally, the probability $\prod$ that a new node will be connected to node $i$ with degree $k_i$ will be

	\[\prod(k_i)=\frac{k_i}{\sum_j{k_j}}
\]

Nodes are added until the size of the graph is as desired. We will start with a graph of $5$ nodes which form a line. We'll add nodes until we have $121$ nodes, with $m=4$. Then we get a graph with  $928$ edges.\\
We now check whether the constructed networks fulfill the necessary conditions. All figures are visible in appendix \ref{cond}. First, we plot the frequency of a degree against the degrees. We see that indeed, in the constructed hierarchical and non-hierarchical network, the higher the degree, the less nodes there are with that degree, while in our random network, the probability of a degree is normally distributed. Next, we plot the cluster coefficient against the degree. The result agrees with our predictions: in the hierarchical network, the cluster coefficient follows a power-law, while in the other networks the cluster coefficient is approximately  the same for all degrees.

\subsection{Evaluating the networks}

One way of evaluating a graph, is by looking at the diameter. The diameter is the biggest distance that exists between two nodes in a graph. The distance between two nodes is the length of the shortest path, that is, the number of edges that minimally have to be passed to go from the one to the other. The shorter the diameter, the easier it is to go from one node to another, so the better the network is. The diameter of the hierarchical network is $2$, the other networks have a diameter of $4$. \\
Now we can look what happens to the diameter of a network when nodes are deleted. There are two ways nodes can be deleted: randomly or delete nodes with the highest degree. These two events are called respectively failure and attack. In a random network, the diameter increases as more nodes are deleted, and there is no difference between failure and attack. In a scale-free network, the diameter remains unchanged if nodes are removed due failure. But if nodes are removed due attack, the diameter increases with a higher scope then in a random network\cite{albert_error_2000}. We let the three constructed networks fail $100$ times, and attack them $50$ times - after this all networks are unconnected. As we expected, in the random network the diameter already increases after a small number of steps ($10$), while in the non-hierarchical and hierarchical network, the diameter remains unchanged during a big number of deletions ($33$ and $68$ respectively). The hierarchical network performs the best, here the diameter remains unchanged until the network falls apart (at step $68$) - see figure \ref{failure} in appendix \ref{f+a}. There is one remark though: the non-hierarchical network has fewer edges then the rest ($928$ instead of $1025$), which could be the cause why it doesn't perform so well. For this reason, we redo the simulation with a non-hierarchical network with $m=5$, which has $1150$ edges. This network performs better, the diameter only changes at step $45$ and the network becomes unconnected at step $69$, but this is still not as well as the hierarchical network - see figure \ref{moreedges} in appendix \ref{f+a}. \\
Attacking the networks shows that indeed in the random network there is not much difference with failure, while the other networks fall apart fast, especially the hierarchical network: the diameter changes immediately, and the network falls apart at step $6$- see figure \ref{attack} in appendix \ref{f+a}.
The phenomenon of civil war could be explained by an attack in an hierarchical network: the node(s) with the highest degree hold the network together, since they have the lowest cluster coefficient. So if such a node is deleted, the network easily falls apart, which causes the leaders of the different clans/clusters to fight for the overall leadership. Civil war is often used as an argument why hierarchical organization is necessary, since if there is no more leader, there is chaos, while this could be caused by the hierarchical structure present from the beginning.


\section{Organization}

A structure isn't enough to simulate real-life systems, most of the time they also have a certain function. We define an organization as a structure with a function. This means the structure as a whole aim at a global goal or a global pattern. It isn't necessary that the goal is clear from the beginning, but there should emerge some general direction. Now, we can define the difference between self-organization and hierarchical organization on the level of the functionality. We speak about self-organization if the function, the global activity, arises spontaneously through local interactions. The common goal is thereby put by the collective. In a hierarchical organization, on the contrary, the structure and function is decided from above, by one or a few agents who put the common goal. 

To be able to reach a goal, some coordination is needed between the different agents. We define coordination as the structuring of actions to minimize friction and maximize synergy\cite{heylighen_self-organization_2011}. There are 4 processes which can help with that: alignment, division of labor, work-flow and aggregation. Next, we will define these concepts, and explain how they can also work in self-organization. In the next chapter, we will make some simulations based on these concepts, and see in which networks there is self-organization, and in which there is hierarchical organization. 

If different agents aim at the same target to avoid friction, we speak about alignment. In a hierarchical organization, this is achieved by all agents adapting towards one leader. In self-organization, agents adapt towards their neighbors, so that a common direction is achieved. This adaptation usually happens by variation and selection: an agent varies a bit in his direction, and the best direction, this is the one with the least friction with his neighbors, is selected. 

Synergy is defined as the surplus gained by working together. A task which couldn't be fulfilled by one individual, can be done by the work of different individuals together. To maximize synergy, first, the initial task is divided into different sub-tasks. Different agents perform different tasks, which is called division of labor. An end product of one work is used for another work, which is called work-flow. Finally, everything needs to be put together, we call this aggregation. This isn't as linear as it looks. At every step in the process it can happen that a task is divided into sub tasks or aggregated with other tasks.\\
At first sight, it may seem that hierarchical organization is needed for this: there needs to be one agent who divides the tasks, who has an overview. But these things can also work by self-organization with simple rules. For division of labor and work-flow, any time an agent doesn't have anything to do, he picks up a task he is most skilled at. An example of this is the evolution of different species: a specie chooses a niche in which he further develops. For aggregation, there are two possibilities: a shared medium or the interaction of the products of different activities. An example of a shared medium is the earth with ant pheromones: ants put pheromones on the earth, other ants can pick up this trace and follow it. An example of the products of different activities that interact, is the ecosystem. The output of one specie is used as a resource or a service for the other specie.\\
The principle behind this is again variation and selection. For division of labor, it works as follows. In the beginning there is already some variation among the agents. This means that some are more skilled for a certain task, so they will select it. By doing the task, they will become even better in it. This is how certain species become extremely good in a certain niche. In the case aggregation happens by different activities that interact, in the beginning there will be some random interactions. Then, the best of these interactions are selected.\\
In the next chapter some simulations of coordination between agents will be presented.

\section{Simulations}

We will now present two simulations of a network of coordinating agents. The basic idea of both models is that agents in the network change their status depending on the status of their neighbors. We simulate this in our 3 different network types, and look at what is different and the same in these networks. In the first model the agents will try to minimize the friction by using alignment, while in the second model they will try to maximize the synergy by using division of labor, work-flow and aggregation.

\subsection{Minimize friction}

In this model, there is a number $n_i$ between $0$ and $1$ assigned to each node $i$, which we can represent by a color on the gray-scale, $0$ being black, $1$ being white. To align, so to aim at the same direction, in this case means to try to have the same color. At each time step, we will therefore update the number of each node towards the number of its neighbors, by the following rule:
\[ n_i \leftarrow n_i + \frac{\sum_{j \in N_i}(n_j-n_i)}{2\cdot |N_i|}
\]
with $N_i$ the set of neighbors of $i$.\\
We assign a fitness function to each agent, based on the idea that the less variation with their neighbors there is (so the less friction), the higher the fitness. 
	\[f(n_i)= \sqrt{\frac{|N_i|}{\sum_{j \in N_i}(n_i-n_j)^2}}
\]
We will also look at the difference in color between a node and the rest of the network. This shows how good a node fits in the overall network. We will call this the colordifference of a node. You could see the colordifference as the inverse of the influence, where we see influence as a measurement of how much agents adapt towards you. Assuming that in the beginning the colors are randomly distributed across the nodes (which is done here), the smaller your colordifference, the more the other agents moved towards your color, thus the bigger your influence. The correspondence is not perfect though, since the colordifference doesn't differentiate whether you were influenced by the other agents- thus your color moved towards their color- or you influenced the other neighbors - thus their colors moved towards your color. 

We will now run this simulation in our 3 networks, and look to how the fitness and the colordifference depends on the degree and the clustering coefficient, in our 3 networks. This is done by plotting the fitness or colordifference of the end of the simulation against the degree or clustering coefficient. 

Plotting the fitness against the degree, shows that in a hierarchical network, the higher the degree, the lower the fitness. In the other networks the fitness is independent of the degree - figure \ref{fitfriction} in appendix \ref{friction}. It seems strange that in a hierarchical network, the nodes with the highest degree, thus the leaders, have the lowest fitness. This can be explained because these nodes or in between different clusters. These clusters evolve rather independently, thus their colors differ a lot. But if we look to the colordifference, we see that in all networks the higher the degree is, the lower the colordifference - figure \ref{colorfriction} in appendix \ref{friction}. Thus, the nodes with a higher degree differ less with the rest of the network then nodes with a lower degree, they are better aligned with the whole network. \\
If we look at the cluster coefficient, we see that in a hierarchical network it behaves opposite to the degree: the higher the cluster coefficient, the higher the fitness and the colordifference (figure \ref{cchfriction} in appendix \ref{friction}). This is logical, since nodes with a lower degree have a higher cluster coefficient. If we look at the non-hierarchical network, we see that the higher the cluster coefficient, the higher the fitness ( figure \ref{ccnhfriction} in appendix \ref{friction}). The explanation is the same from above: the lower your cluster coefficient, the more your neighbors or disconnected, thus the more they will evolve independently, thereby the more difficult it is to adapt to them, thus the lower the fitness. The colordifference is independent of the cluster coefficient in a non-hierarchical network - figure \ref{ccnhfriction} . In a random network, both the fitness and the colordifference are independent of the cluster coefficient (figure \ref{ccrfriction} in appendix \ref{friction}).\\
There are two basic mechanisms in place in all networks we can extract from this analysis. The higher the cluster coefficient, the higher the fitness. And the higher the degree, the lower the colordifference is. If the cluster coefficient and the degree are uncorrelated, the rest of the combinations are uncorrelated. Note that in a random network the fitness is independent of the cluster coefficient, but this could simply be because there is not enough difference in cluster coefficients of different nodes, maybe the fitness only increases as soon as a certain threshold is passed.

We'll now give the mean and standard deviation of both the fitness and the colordifference in all networks, and discuss these numbers. 
The mean fitness is $131.7$, $124.4$ and $92.0$ for respectively hierarchical,
non-hierarchical and random network, while the standard deviation is respectively  $60.3$, $38.1$ and $21.4$. Thus, the fitness in the hierarchical network is a bit higher then in a non-hierarchical network, but there is more difference among the nodes. The random network is particularly worse then the other networks, even tho there is less difference. \\
 The mean colordifference is respectively $0.0353$, $0.0126$ and $0.0078$ for hierarchical,
non-hierarchical and random network, and the standard deviation is respectively $0.0090$, $0.0048$ and $0.0039$. Thus, the difference between the node colors of a hierarchical network is bigger then in the other networks, which means that in general there is less alignment - nodes are aligned with there neighbors but not with the whole network. There is also more variation of the colordifference between the nodes of a hierarchical network.   
\subsection{Maximize synergy}

The next model is inspired by the ecosystem. Each agent needs certain products and produces products others can use. We can represent this by assigning a vector of length $n$ to each node, with $n$ the number of products in the system. We call this vector the needvector $\bar{n}_i$ of node $i$. If a node needs a certain product, we put a $1$ on that place, if not, we put a $0$. In the beginning, we put $m$ ones on a random place in each needvector. 
For each system a productionvector $\bar{p}$ is also created. Each product an agent has, is changed by that agent into another product, a garbage product. This is done by a random permutation of the products, represented by the productionvector. This is a vector containing the numbers $1$ until $n$, but in another rank. Product $k$ becomes the product on position $k$ in the productionvector, this is $\bar{p}(k)$. This productionvector is the same for all agents.

But which products are transformed by an agent? This is the food an agent $i$ has, we'll represent this in its foodvector $\bar{f}_i$, and we'll present two models to construct this vector. In the first model, a non-propagation model, every product in the needvector of node $i$ is a food product, thus $\bar{f}_i=\bar{n}_i$. In the second model, where there is propagation of food, only the food an agent has obtained is produced into a garbage product, which can be used by its neighbors. In the beginning, an agent has all the food products from his needvector, thus the two vectors are the same. In each next step, an agent obtains all the food products he needs and one of its neighbors has. The rest remains the same for both models. So at step $1$ $\bar{f}_i=\bar{n}_i$, but from step $2$ on, the vector $\bar{f}_i$ of a node is generated by:

\begin{equation}
	\mbox{for } k=1...n \mbox{    } \ \ \bar{f}_i(k)= \left\{ \begin{array}{ll}
	     1 & \mbox{ if } \bar{n}_i(k)=1 \mbox{ and } \exists j \in N(i): \bar{g}_j(k)=1 \\
	     0 & \mbox{ else } 
	     \end{array} \right. \label{f_i}
	     \end{equation}
with $N(i)$ the set of neighbors of $i$, and $\bar{g}_j$ the garbagevector of node $j$ as constructed below.
             
These food products are then transformed into garbage products by the productionvector. The garbage products are available for its neighbors. Thus, for each node $i$ we can define a garbage vector $\bar{g}_i$ depending on the foodvector $\bar{f}_i$ by :

\begin{equation}
	\bar{g}_i(\bar{p}(k))=\bar{f}_i(k) \ \ \mbox{ for } k=1...n. \label{g_i}
\end{equation}                    

The fitness of an agent $i$ is the sum of all the garbage products one of its neighbors has and he needs. 
\begin{equation}
	 \mbox{fit}(i)=\sum_{k|\bar{n}_i(k)=1 \ \wedge \ j\in N(i)}{\bar{g}_j(k)}. \label{fit}
\end{equation}
In each step, the needvector of each agent is changed by a variation and selection, based on the genetic algorithm\cite{holland_genetic_1992}. We construct $10$ random mutations of the needvector, and choose the one with the highest fitness. We create a mutant by choosing two different random positions in the needvector and flip places. Thus, if we have chosen two numbers $k_1$ and $k_2$ between $1$ and $n$ randomly, and the needvector is
	\[\bar{n}_i=(\bar{n}_i(1),...,\bar{n}_i(k_1),...,\bar{n}_i(k_2),...,\bar{n}_i(n))
\]
then our mutant is
	\[\bar{mut}_i= (\bar{n}_i(1),...,\bar{n}_i(k_2),...,\bar{n}_i(k_1),...,\bar{n}_i(n)).
\]

$10$ mutants are created, and the one with the highest fitness replaces the original needvector (if this fitness is higher then the original). This assures that the fitness of an agent will rise. In the propagation model, he will use more garbage of its neighbors (by definition of the fitness), hence more food, therefore he will also produce more garbage for its neighbors. Thus the fitness of its neighbors will also increase, and this effect will spread across the network. In the non-propagation model, the amount of garbage produced isn't influenced by a better needvector, thus the increase in fitness doesn't get propagated. To see how this really works, we plot the evolution of the fitness in time, in all networks, both in the non-propagation model and the propagation model. In the non-propagation model, in the beginning the fitness increases quite fast, but it quickly converges towards one optimum (see figure \ref{nopropfitt} in appendix \ref{fitt}). In the propagation model on the other hand, while in general the scope decreases over time, at time step $3$ the scope increases, and the fitness is still increasing at the end (see figure \ref{propfitt} in appendix \ref{fitt}). The fitness is also significantly bigger in the propagation model.  This shows the fitness gets propagated through the network in the propagation network. While the values are different in the different networks, the fitness evolves in the same way. \\

To see how the need- and garbagevector gets propagated through the network, we first look at a system where $\bar{p}=(1,...,n)$, thus food products are transformed into themselves, $\bar{g}_i=\bar{f}_i \ \forall i$. For the non-propagation model, this means need-, food-, and garbagevector are the same, thus there is only one vector per agent. An agent will try to adjust his vector to that of his neighbors (he'll adjust his needvector to the garbagevector of its neighbors, since that results in the most fitness). By this mechanism, vectors get propagated. In the propagation model, an agent will still strive to have as much garbage products of its neighbors as possible, thus to equalize its needvector with the garbagevectors of its neighbors. Since $\bar{g}_i=\bar{f}_i$, it will thereby strive for a needvector which is as equal as possible with the needvector of its neighbors, especially on the positions where the needvector is equal with a neighbor of this neighbor - since then there is more chance the need will be fulfilled, and thus on this position there will be a food product. Seen from another point of view, an agent will propagate its needvector on the positions where its need is fulfilled, by sending the garbagevector to its neighbors ($\bar{g}_i=\bar{f}_i$). These neighbors will evolve their needvectors towards this vector, and by this propagate it further. \\
In the general model where $\bar{p}$ could be any permutation vector, the basic mechanism is still the same, only it is no longer true that $\bar{g}_i=\bar{f}_i$. Thus, at each agent, the vector first get transformed before it is propagated further. An agent evolves his needvector towards the garbagevector of its neighbors, transforms this vector into the foodvector and then the garbagevector, and propagates this vector further, where his neighbors do the same thing. Thus the needvectors won't align towards each other anymore, but they will evolve so that the garbage in the system is used as much as possible.\\ 

To summarize the propagation model, the following steps are done for each agent $i$: 
\begin{enumerate}
	\item construct $\bar{f}_i$ by \eqref{f_i}
  \item construct $\bar{g}_i$  by \eqref{g_i}
  \item compute fitness by \eqref{fit}
  \item transform $\bar{n}_i$ by variation and selection  
\end{enumerate}
The non-propagation model is the same, except that the first step isn't done, $\bar{f}_i=\bar{n}_i$.\\

  We now plot the difference in fitness against the degree and the cluster coefficient at the end of the simulation. With the difference in fitness we mean the difference in fitness of a node between the end and the beginning of the simulation. This shows how much a node got stronger during the simulation. We could also just look at the fitness, this gives the same results, thus in the beginning the fitness is approximately the same for all nodes.\\
 In all three networks, for both models, we see that the higher the degree, the higher the difference in fitness, and the higher the cluster coefficient, the lower the difference in fitness - see appendix \ref{synergy}. Thus, nodes with a higher degree or a lower cluster coefficient will increase more in fitness. In the non-propagation model, this effect is the highest in a hierarchical network, and only a little bit there in a random network. For the cluster coefficient, it is also in a non-hierarchical network that a higher cluster coefficient only leads to a little bit lower difference in fitness.\\
We now look at some numbers. This is first done for our first model, where there is no propagation. The mean difference in fitness is $3.21$, $3.17$ and $3.07$ for respectively hierarchical, non-hierarchical and random network. Thus the random network performs worse than the other networks.\\
The standard deviation of the difference in fitness is $3.15$, $2.58$ and $2.24$ for
respectively hierarchical, non-hierarchical and random network. Thus there is more variation in a hierarchical network.\\
For the model with propagation, the mean difference in fitness is respectively $10.4959$, $7.8595$ and $7.0083$ for the hierarchical, non-hierarchical and random network, while the standard deviation is respectively $16.5817$, $6.7297$ and $2.9958$. Thus the same mechanism as in the non-propagation model is working, but especially for the standard deviation there is more difference between the networks. This model performs better then the model without propagation - the mean difference in fitness is higher - even though the standard deviation is also higher, thus there is more difference between the nodes.

What also looks interesting to investigate, is how the difference in color or food between the end and the beginning, depends on the degree or the cluster coefficient. This variable shows how much a node changes during the simulation, thus how much it is influenced. We see though that in both models the difference in color or food between the end and the beginning is independent of the degree and the cluster coefficient.

 \section{Further research}

There are many ways the last model can be updated. The first way is to use another fitness function. Since in the above model we work only binary - either an agent has fulfilled his need for a product and thus produces it into a garbage product, or he hasn't - it would be more logical to count each product only once.  Thus the fitness function of a node $i$ would look like this
\begin{eqnarray}
\mbox{fit}(i)&=& \sum_{k|n_i(k)=1}{\max_{j\in N(i)}{g(j,k)}} \nonumber \\
    &=& \sum_k{f(k)}\nonumber
\end{eqnarray}
A function to measure the influence of agents in the second model can be practical. This can be the amount of garbage used by the neighbors of an agent. The higher this number, the more the neighbors depend on that agent. Or it could be the garbage given divided by the garbage received. Or it could be a measurement how much the agent has changed. If he has changed a lot, he had to adapt a lot to its neighbors, if not its neighbors where probably more adapting themselves towards the agent.  

The idea to try to minimize the friction can also be introduced. In this model this could mean that it's bad for your fitness if other agents also want a product you want, and good if more neighbors have a product. This can be modeled by allowing any positive number in the food- or garbage vector, not just $0$ or $1$. An agent divides his garbage equally over the neighbors who need it. Thus the amount of a certain food product an agent has is the sum of the amount of garbage he receives from his neighbors. In this model, the fitness will be computed as the sum of all the food a node has received, thus $\sum_k{f(k)}$.

The network structure itself can also be changed. This can be done by removing and adding edges. A node can connect with neighbors of neighbors, with more chance if this increases his fitness more. On the other hand a node could remove a connection if that neighbor (almost) don't provoke any rise in fitness. Maybe the amount of connections could be constrained. 

Some biological and genetic ideas can also be added. The evolution mechanism of the needvector can be made more complex by using the genetic algorithm more\cite{holland_genetic_1992}. A partner is chosen by fitness out of the group of neighbors with enough in common. Then a cross-over and mutation happens. \\
The notions of birth and death can also be included, by adding and deleting nodes. A child can be created by the cross-over and mutation mechanism, and connected with his parents. First, he gets the same fitness as his parents, he gets mother milk or is fed by his parents, thus he don't need to fulfill his own needs yet. As times goes by, he gets connected with more agents (by the mechanism described above), learns to be more independent, and computes his own fitness. An agent can die in two way. First, out of hunger, thus the lower the fitness the more chance to die. Second, by being eaten by another agent. This can be modeled by making some garbage vital. The bigger the difference is between the attacker's fitness and the prey's fitness, the more chance the prey get killed (and thus deleted from the network).\\
We can also use this model to try to model the evolution of life, how things become more and more complex. The further a product is in the vector, the more complex it is. The first product is produced into the second product, that product into the the third, and so on. Or, a more realistic model: the 1st and the 2nd product are produced into the 3rd, the 2nd and 3rd into the 4th and so on. Thus, $(12)\rightarrow 3,(23)\rightarrow 4, (34)\rightarrow 5,...$ The first (or first two) products are always available, this could be sunlight. 

We could allow transporters or men in the middle: an agent who doesn't process products, but immediately pass it over to its neighbors. Thus (some of) his food products will be the same as his garbage products. The man in the middle can be seen as a catalyst in chemistry: he doesn't do anything in the process, but makes the reaction possible between two other agents. The question is now whether it is needed to give some benefit to the man in the middle to transport the goods. A catalyst doesn't have any benefit, and actually it is better to bypass the man in the middle and have direct contacts, since that's more efficient. The benefit exist if agents strive for connections which increase their fitness (thus which give them garbage). You can also adapt the fitness function so that giving garbage also leads to an increase in fitness. 

Some of these ideas can be used to build a model of an alternative economy. The idea is that an agent has a need vector which is constant, and a production vector which changes due variation and selection. As before, the agents form a network and the fitness is the sum of the fulfilled needs.
Production happens if the resources coded in the production vector are available. The amount of end product produced is the amount of resources multiplied with some efficiency variable. This variable increases logarithmic as there is more produced in this manner. In the beginning an agent isn't good at it yet, thus only a little bit is produced. But he learns fast how to do it. Once he knows how to do it, he won't get that much better in it. 
The way products are used follows $4$ steps. First, all agents use the products they have to fulfill their needs, directly or by first going trough a production process. Next, they give their products away to the agents who need it directly or who can produce products they need with it. Third, they give it away to the agents who use it in their production process. Finally, they divide the products that are still left (if any) to all their neighbors, who just transport it. \\
The economy is organized - besides the evolution of the production vector - by the links created and deleted between the agents. If $A$ is connected to $B$ and $B$ to $C$, then $A$ gets connected to $C$, with a higher chance if more needs are fulfilled. The amount of edges of an agent is limited, thus connections with certain agents are deleted. If a neighbor doesn't supply a lot of products the agent need but takes a lot of products, he doesn't add much for the agent, and thus the connection is deleted. This is a way to avoid the development of parasites. Even without having a direct profit form giving away products, it is still stimulated. 
The details of this model still needs to be worked out.  

Other structures then graphs can be used. This can be a directed graph, hierarchy could then be defined as some kind of order relation. We could also work with a hypergraph, where edges can consist of more then two nodes. Here edges can be seen as a market, or for our model more as a free shop, where products can be dropped and taken. The next step is to look what these structures do in the discussed models.

\section{Conclusion}

We've made some interesting simulations which give better insights in how global organization arise from local interactions. In neither of the networks we found agents who had all or a lot of influence. Thus we didn't saw a hierarchical organization, but maybe this is because influence is still poorly defined. They are some differences in behavior in different networks though. The random network performs particularly worse then the rest, and the hierarchical network performs the best. But there is more inequality, more variation in fitness, in the hierarchical network, and the friction in the whole network is also higher: mean colordifference is higher and more vulnerable to attack. There is also more dependency on the degree and the cluster coefficient (both for the fitness and the colordifference).

\bibliographystyle{plain}
\bibliography{RefpaperSOvsHO}




\appendix

\newpage
\section{Appendices}
\subsection{Necessary conditions 3 networks}\label{cond}

\begin{figure}[htbf]
\begin{center}
\includegraphics[scale=0.3]{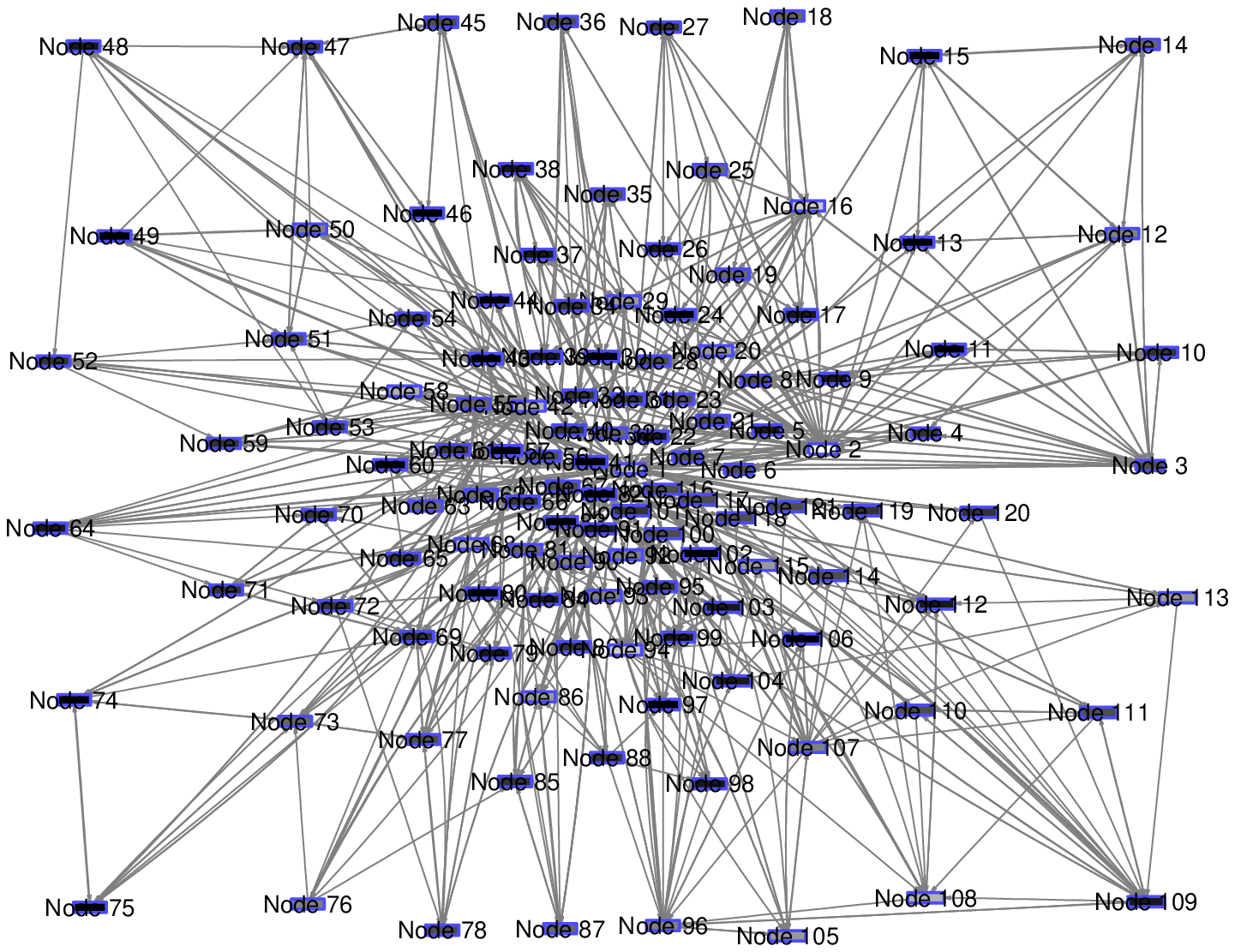}
\includegraphics[scale=0.3]{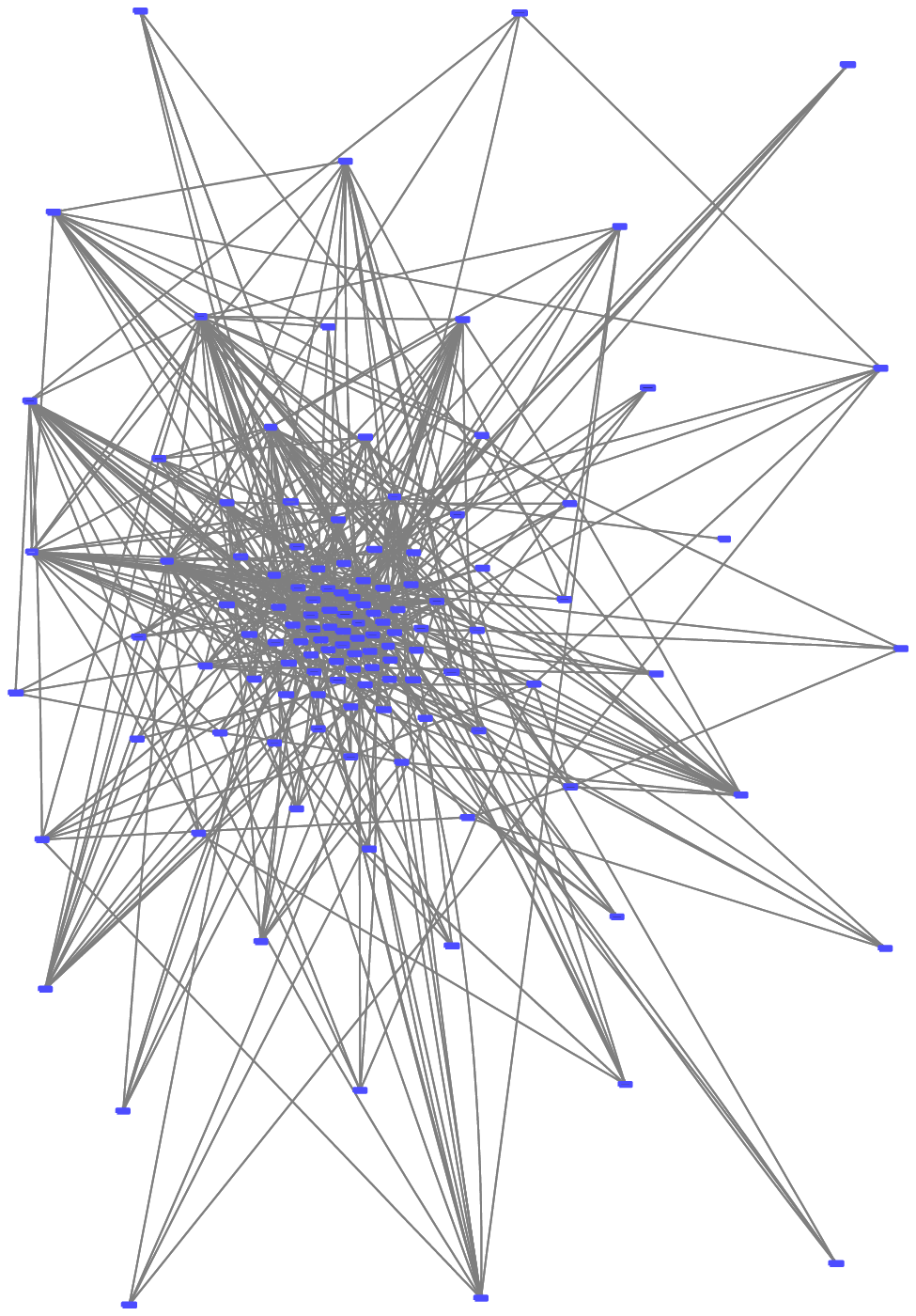}
\includegraphics[scale=0.3]{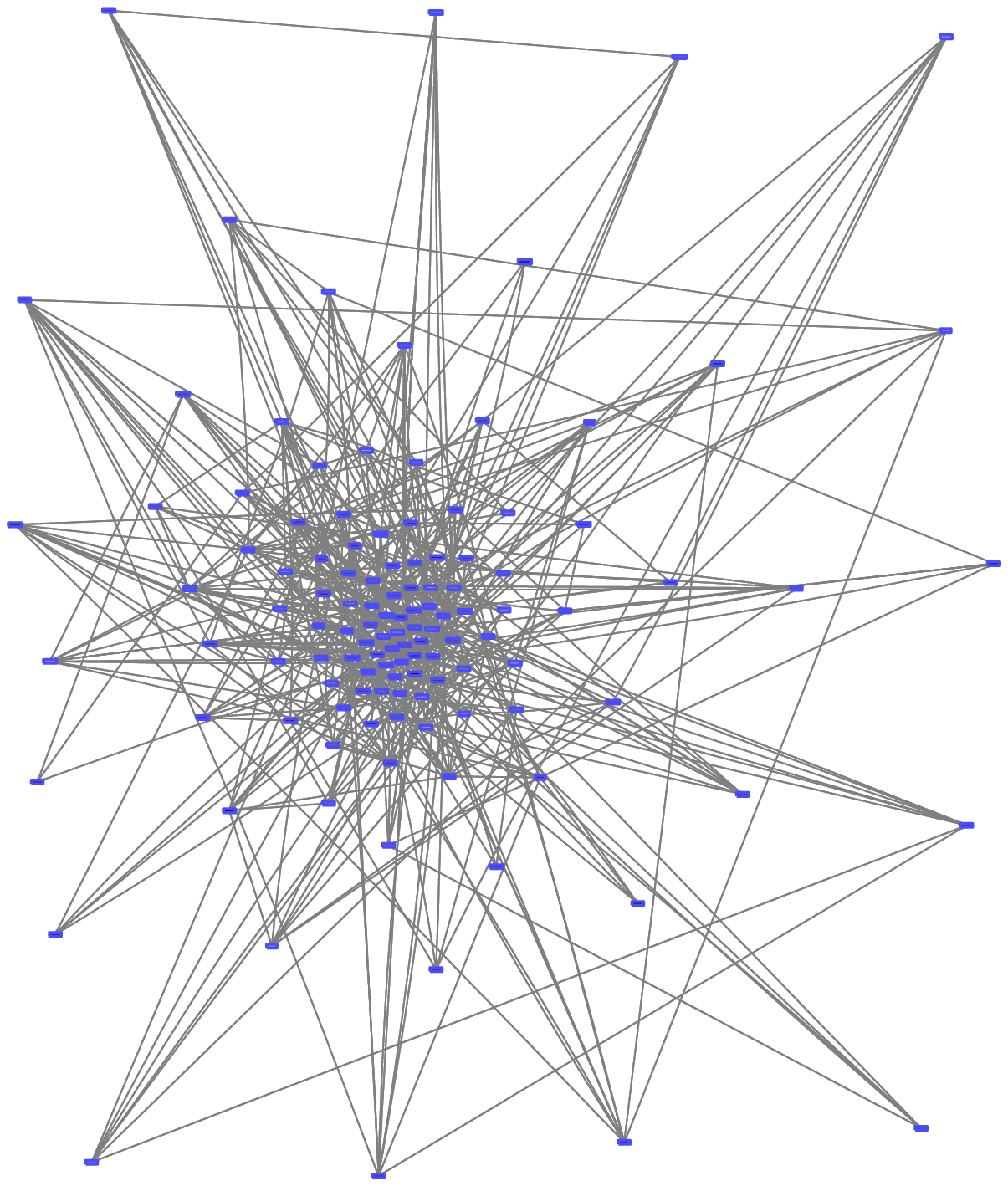}
\caption{Hierarchical, non-hierarchical and random network}
\end{center}
\end{figure}

\begin{figure}[htbf]
\begin{center}
\includegraphics[scale=0.3]{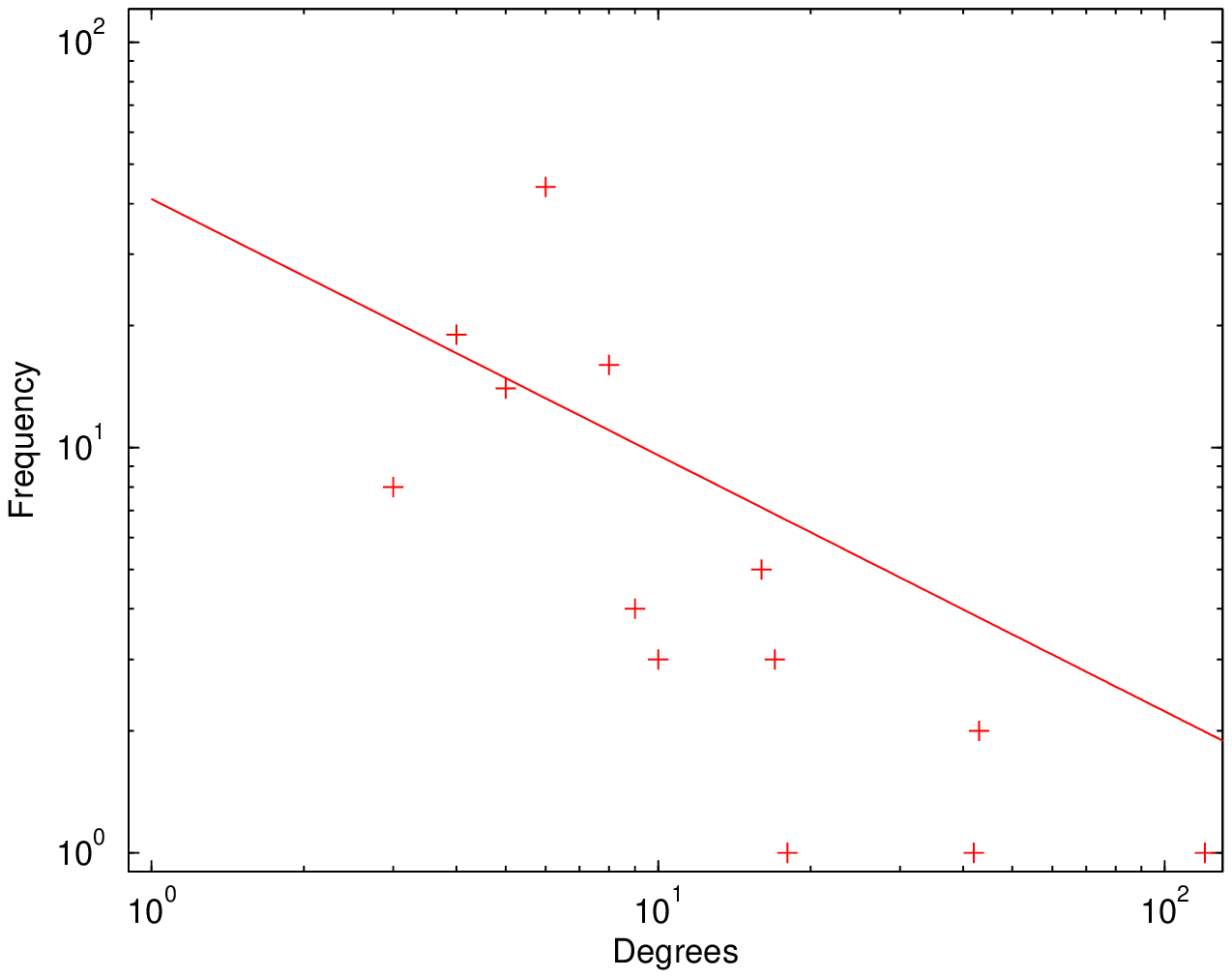}
\includegraphics[scale=0.3]{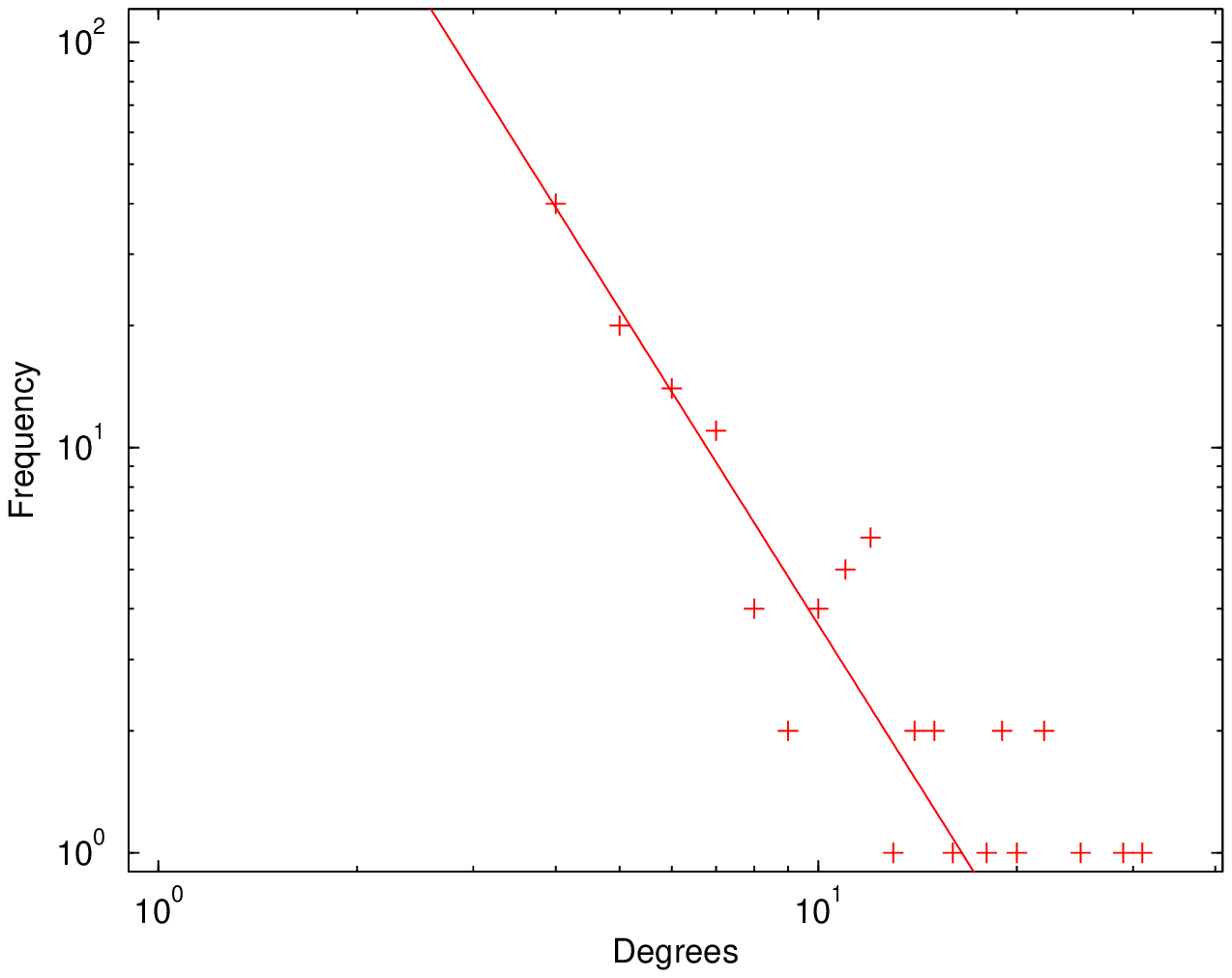}
\includegraphics[scale=0.3]{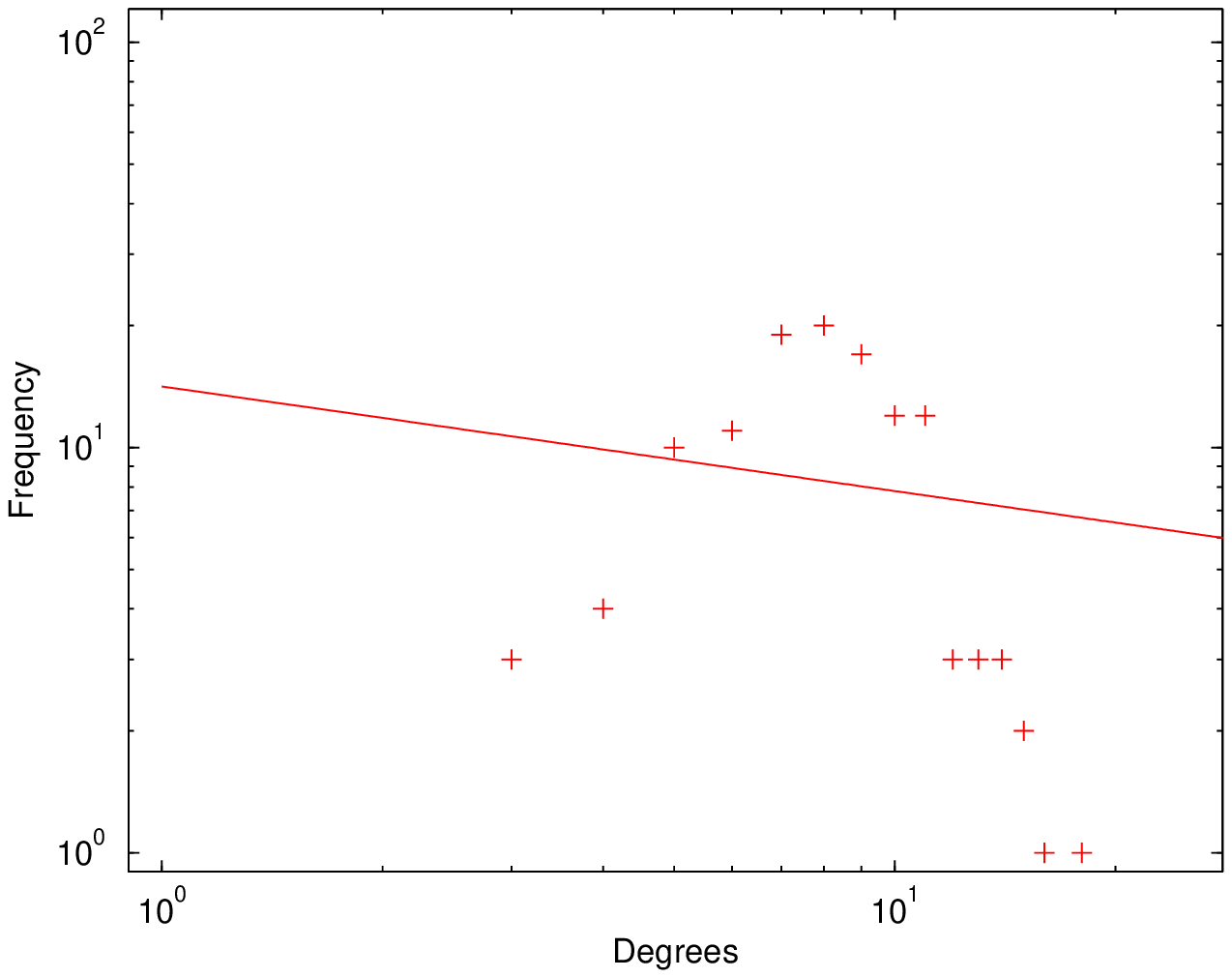}
\caption{Plot probability distribution of resp hierarchical, non-hierarchical and random network}
\end{center}
\end{figure}

\begin{figure}[htbf]
\begin{center}
\includegraphics[scale=0.3]{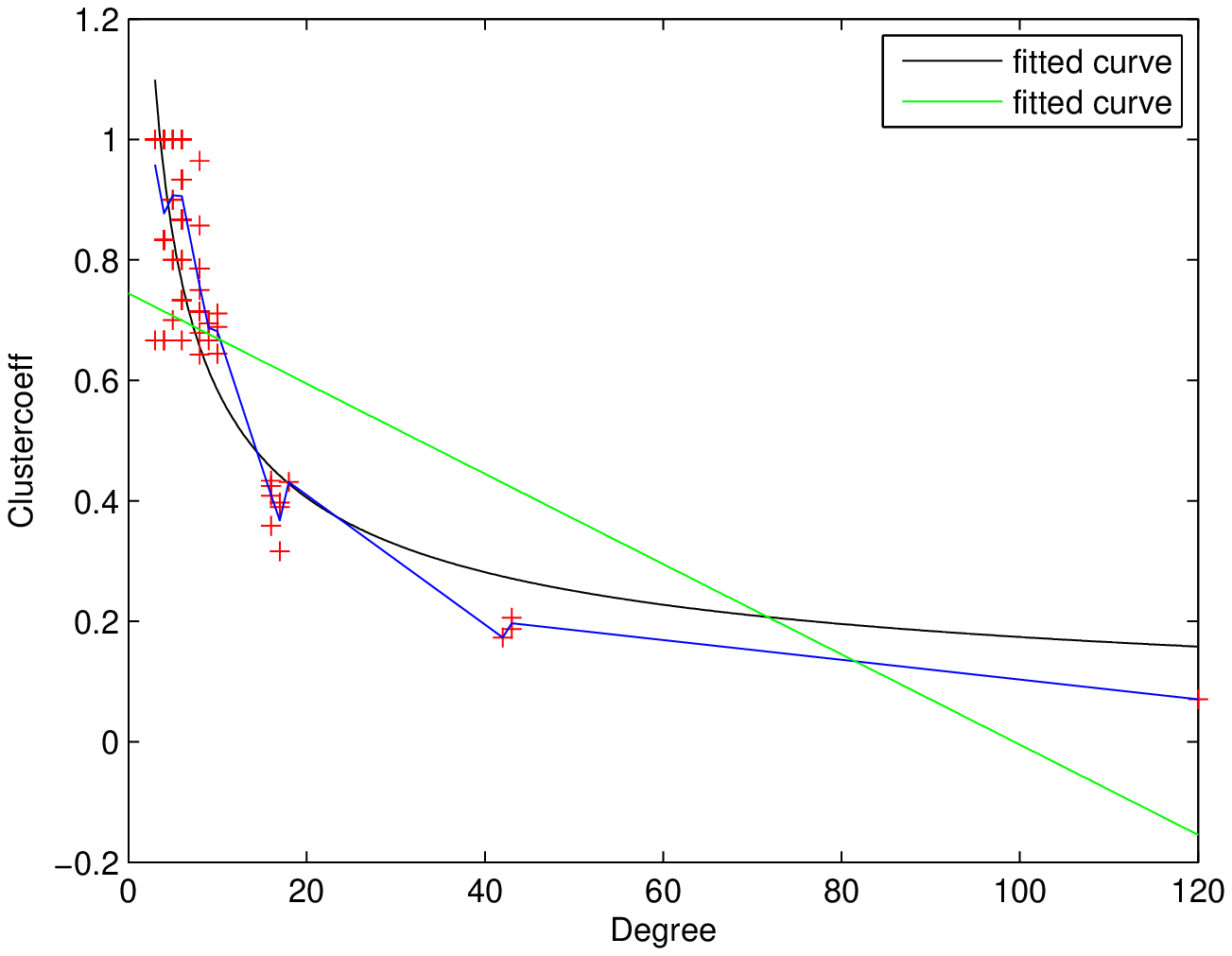}
\includegraphics[scale=0.3]{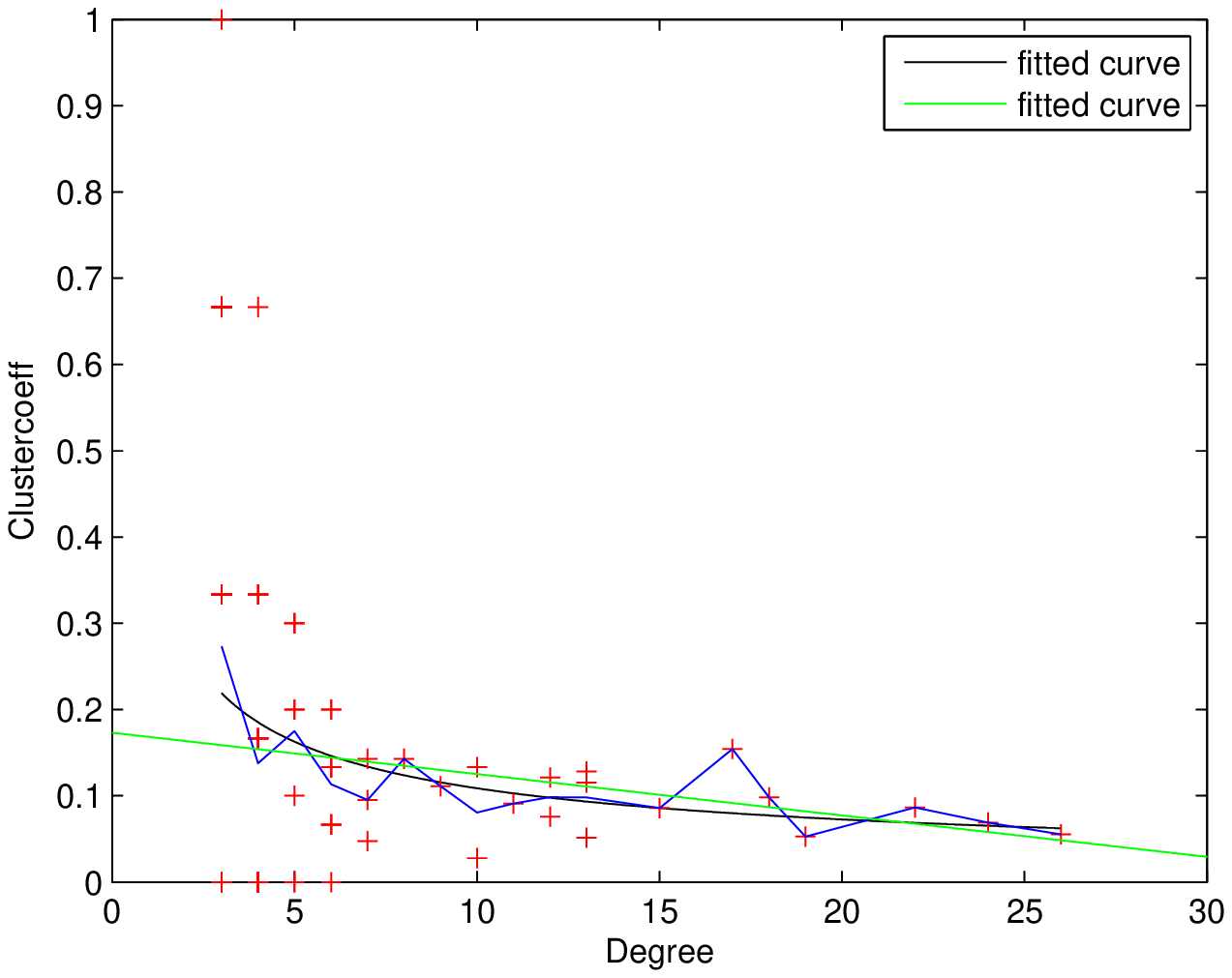}
\includegraphics[scale=0.3]{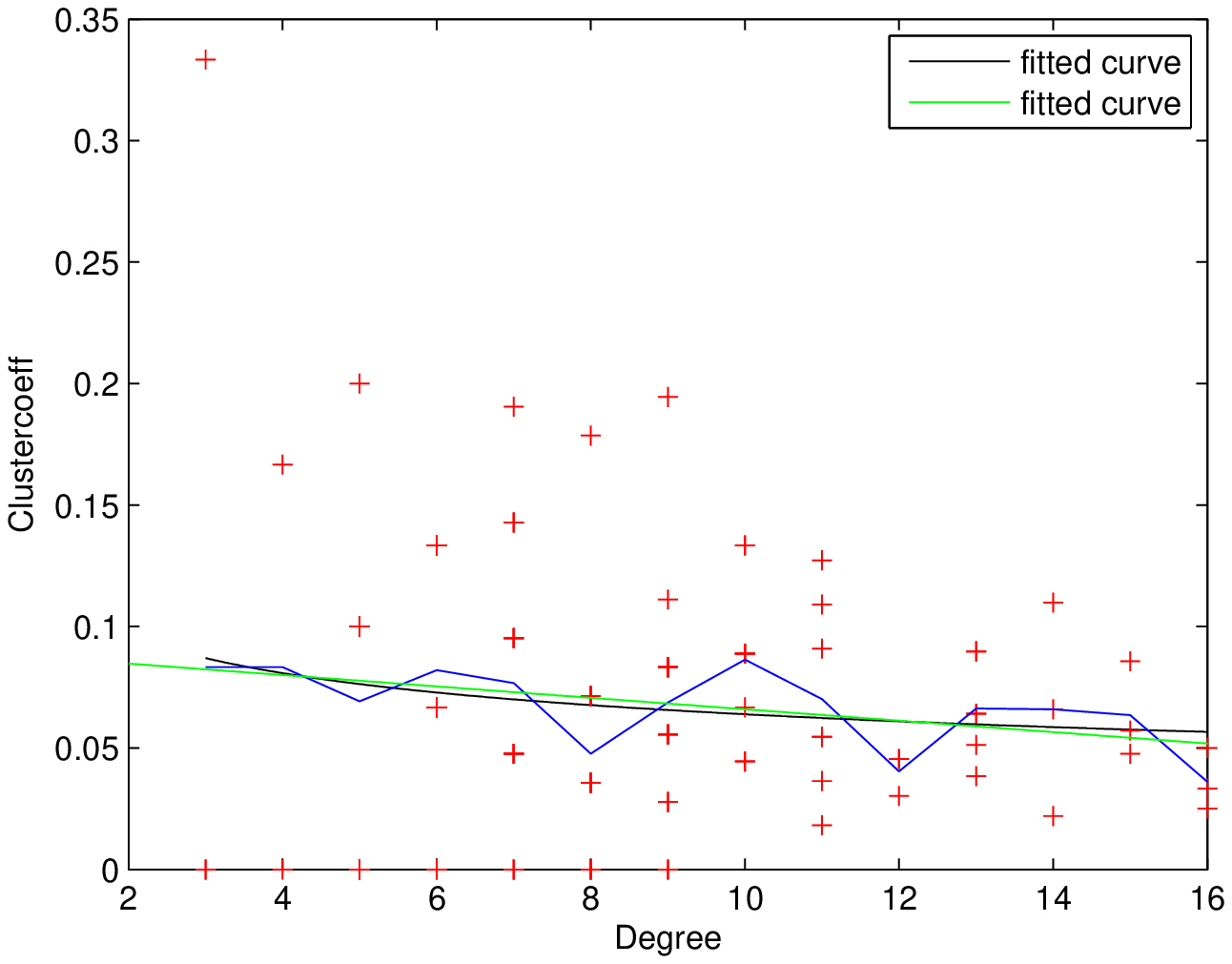}
\caption{Cluster coefficient against degree in resp hierarchical, non-hierarchical and random network}
\end{center}
\end{figure}

\newpage
\subsection{Failure and attack}\label{f+a}

\begin{figure}[htbf]
\begin{center}
\includegraphics[scale=0.3]{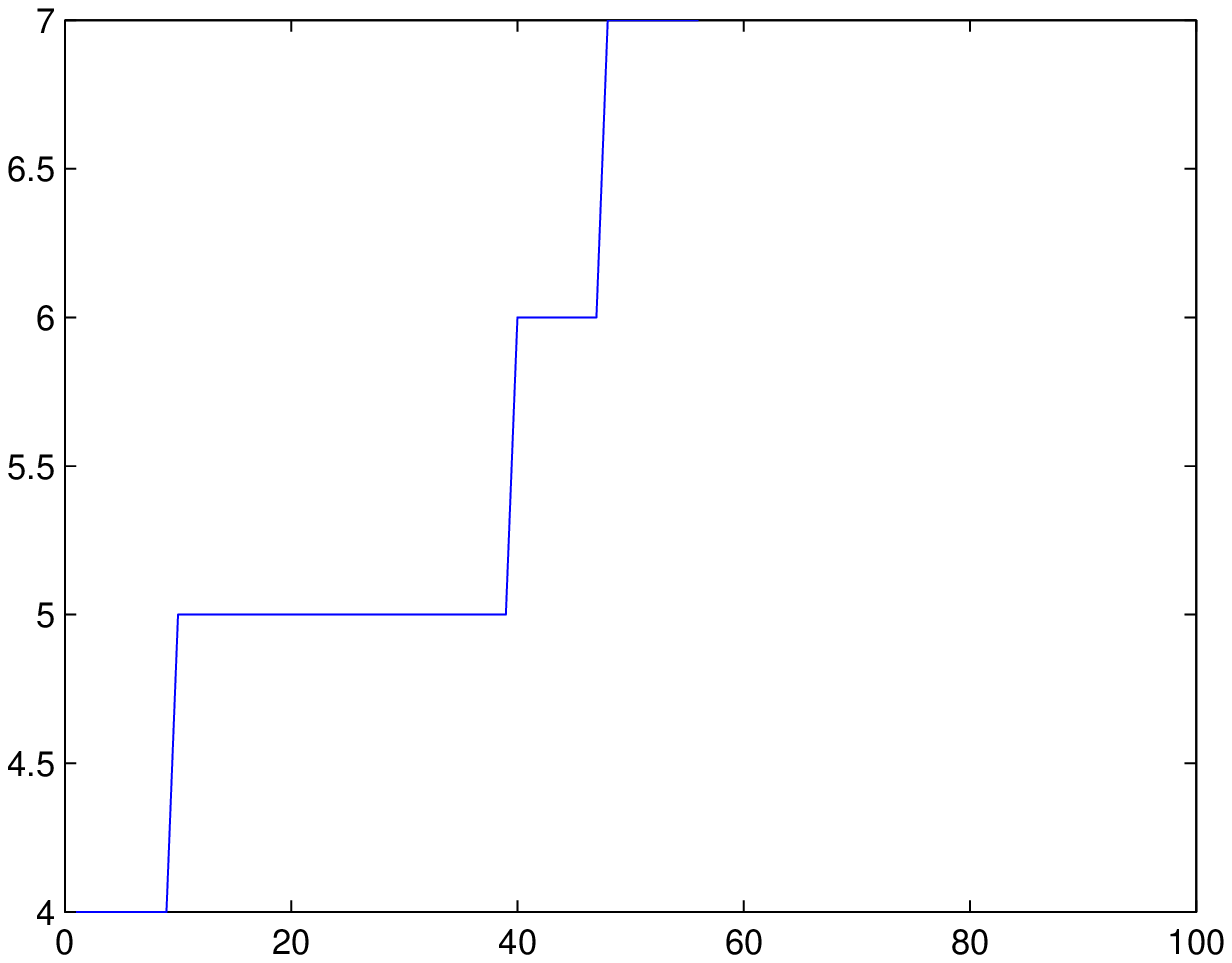}
\includegraphics[scale=0.3]{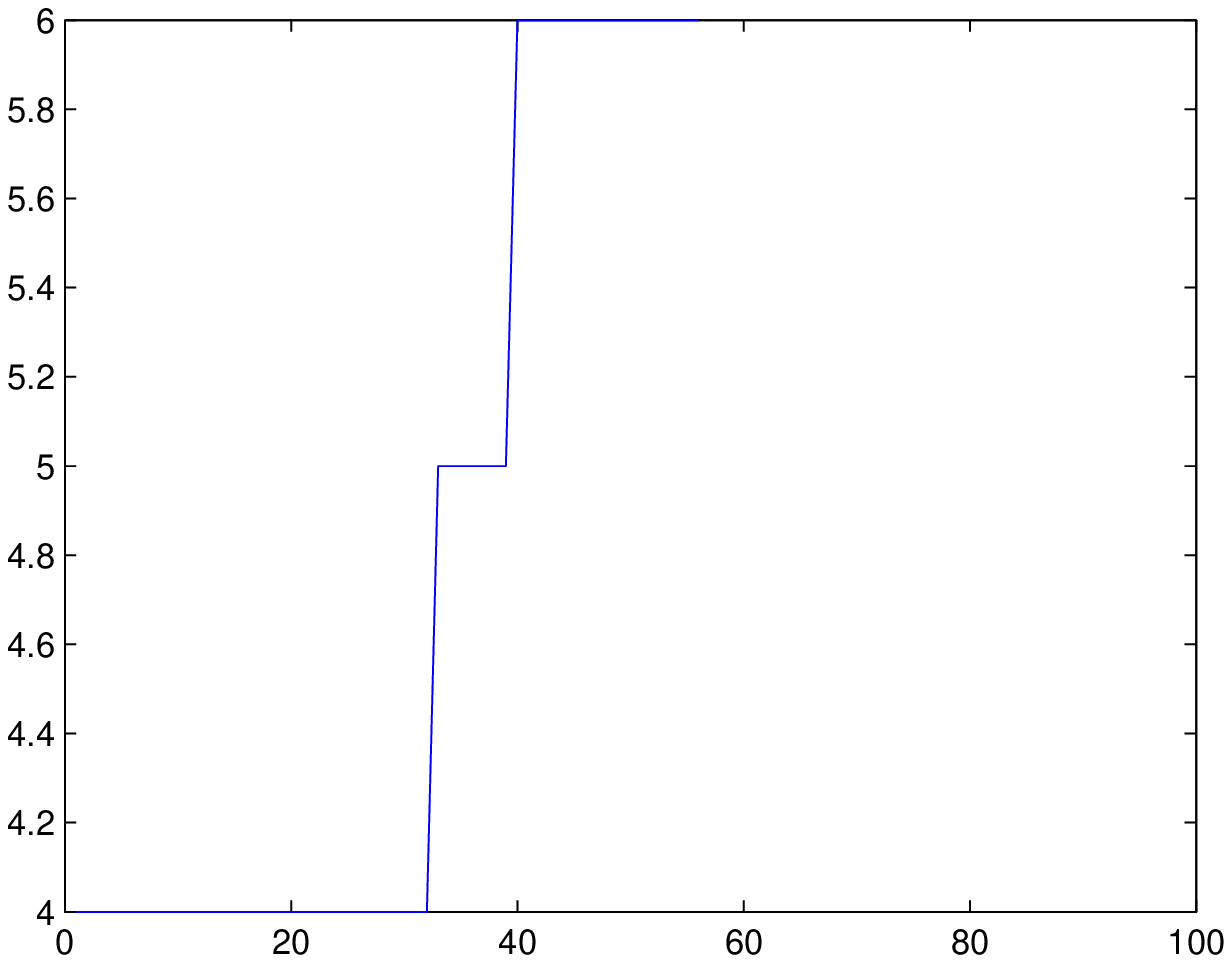}
\includegraphics[scale=0.3]{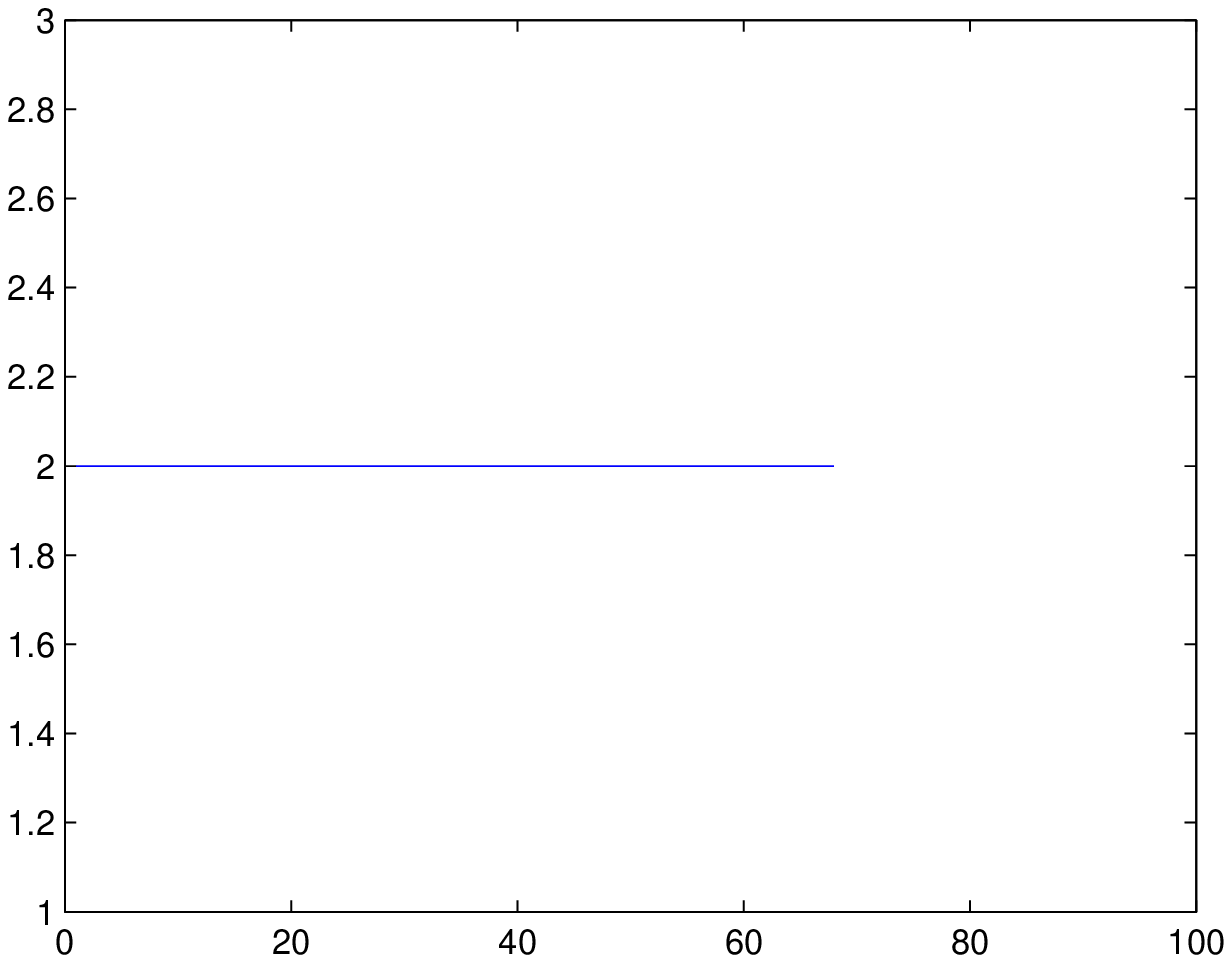}
\caption{Failure in resp random, non-hierarchical and hierarchical network}\label{failure}
\end{center}
\end{figure}

\newpage

\begin{figure}[htbf]
\begin{center}
\includegraphics[scale=0.4]{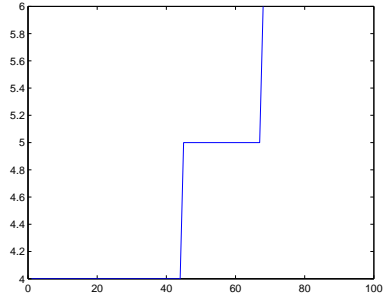}
\caption{Failure in a non-hierarchical network with $m=5$, thus more edges}\label{moreedges}
\end{center}
\end{figure}

\begin{figure}[htbf]
\begin{center}
\includegraphics[scale=0.3]{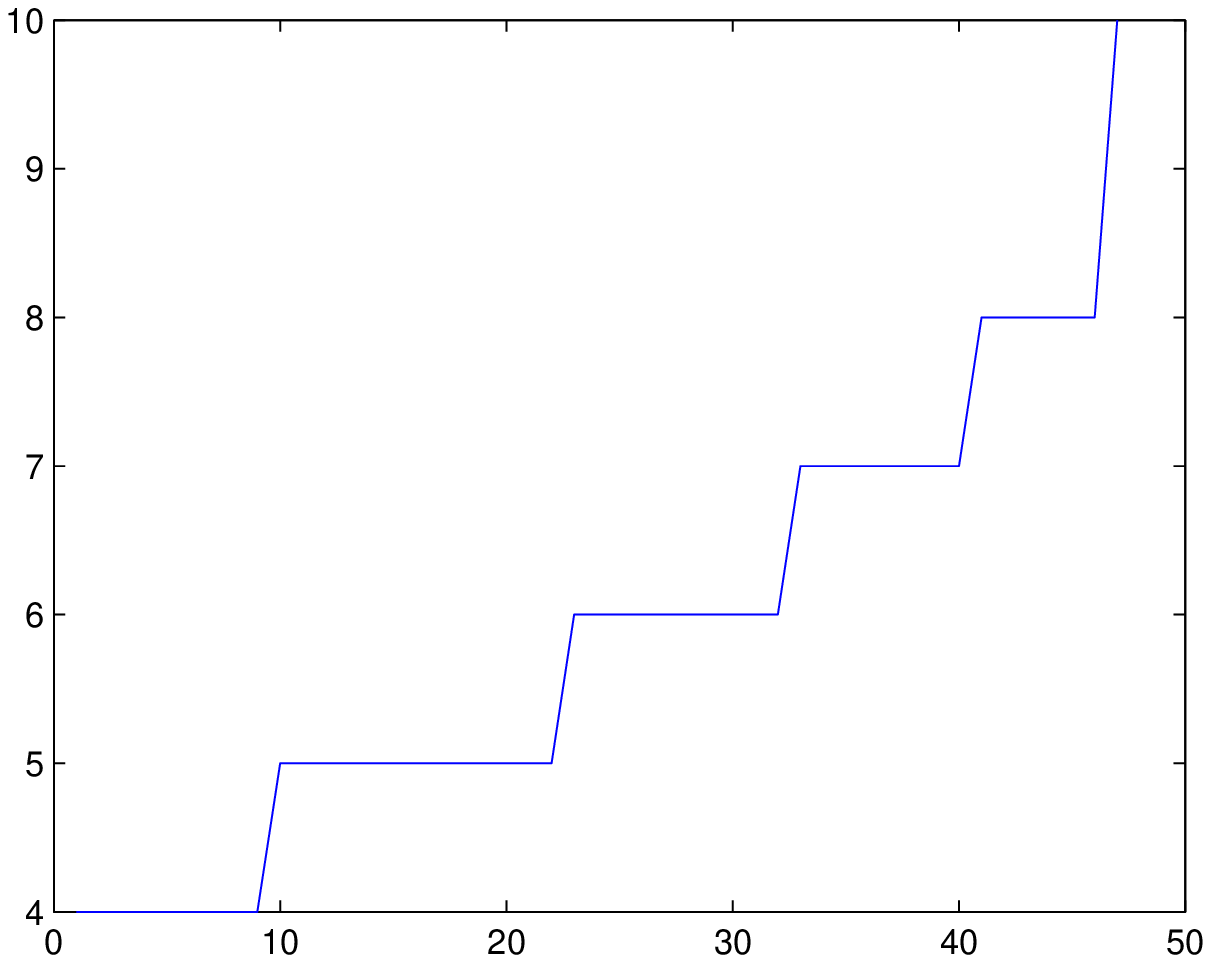}
\includegraphics[scale=0.3]{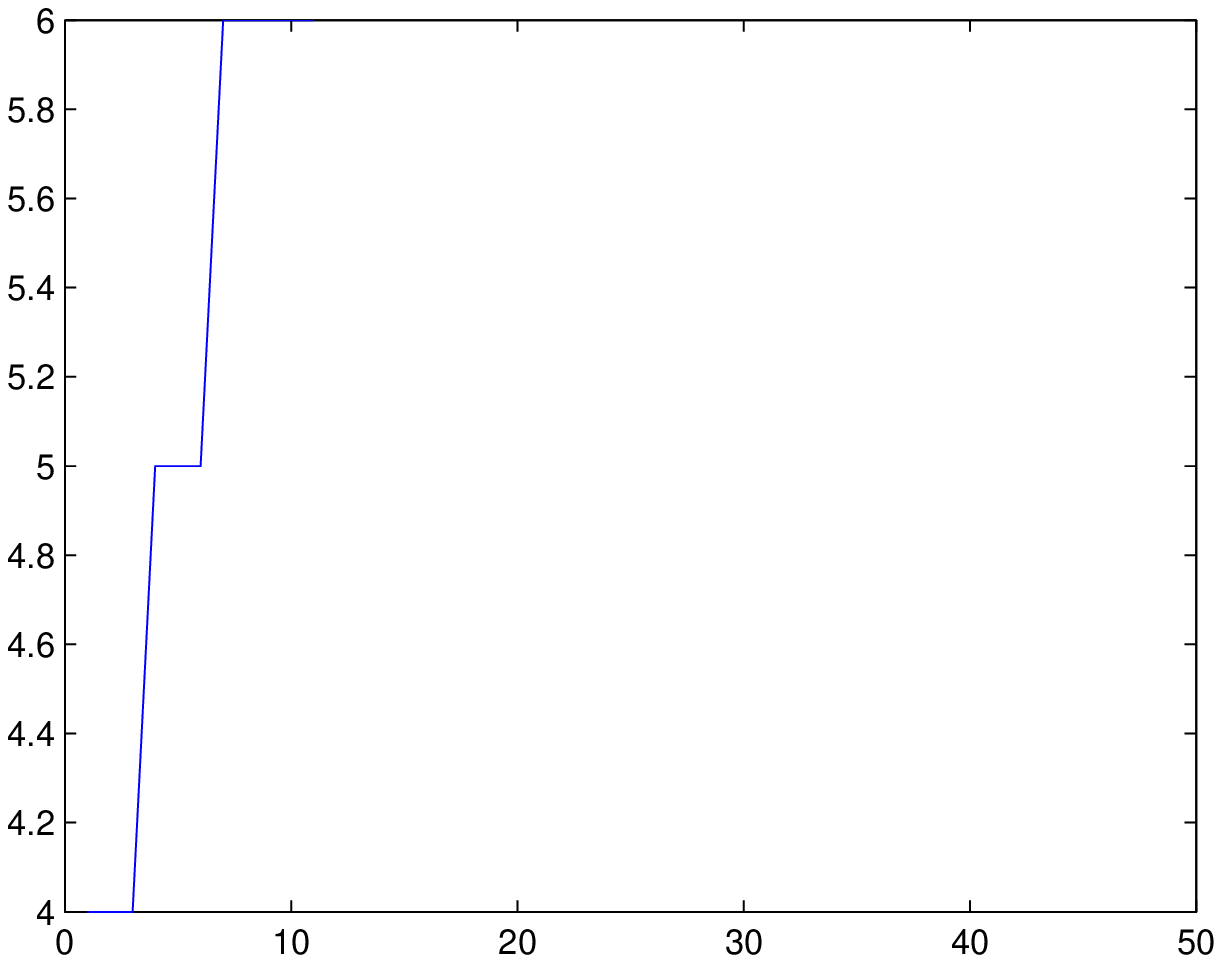}
\includegraphics[scale=0.3]{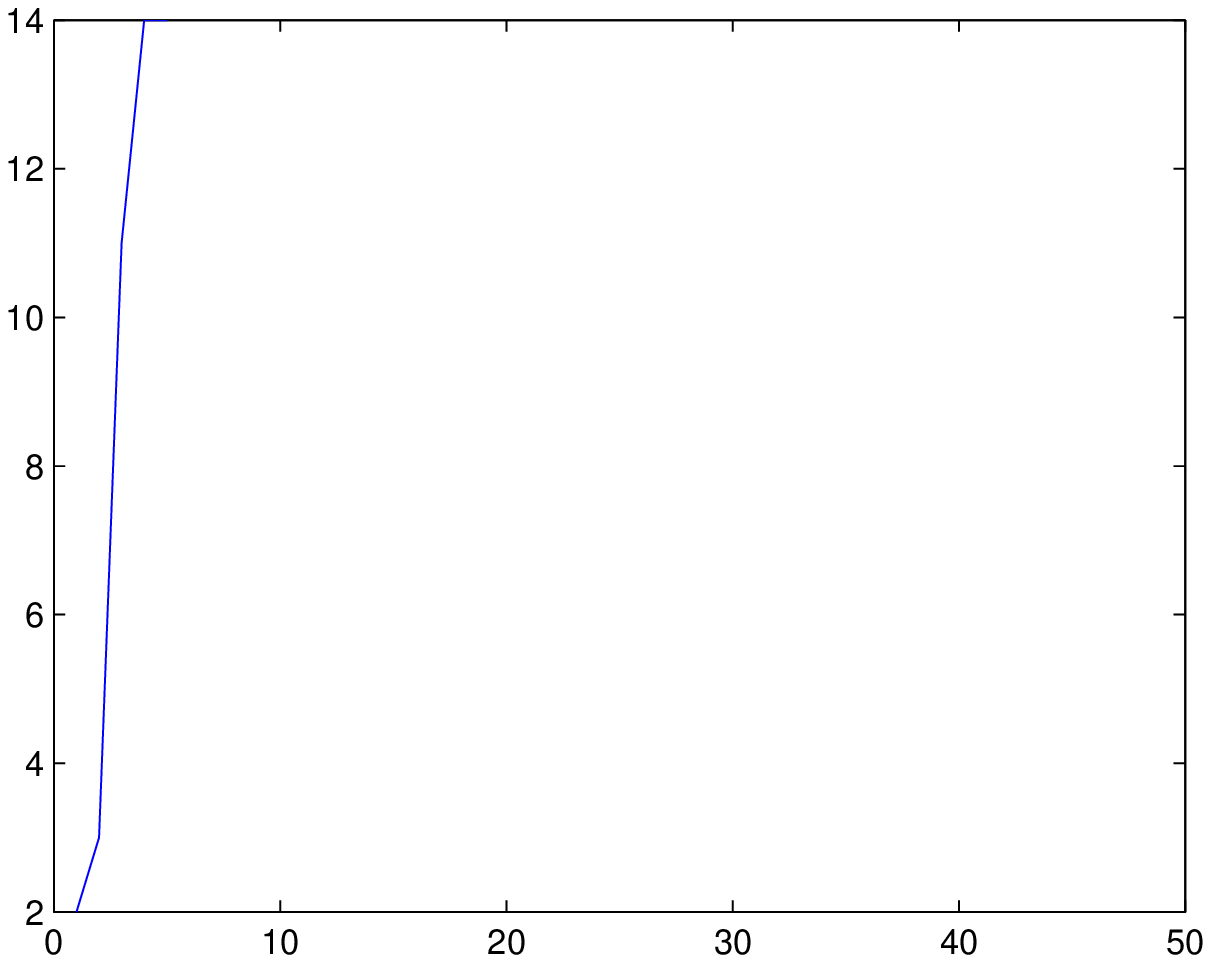}
\caption{Attack in resp random, non-hierarchical and hierarchical network}\label{attack}
\end{center}
\end{figure}

\subsection{Plots minimize friction model}\label{friction}
\newpage

\begin{figure}[htbf]
\begin{center}
\includegraphics[scale=0.3]{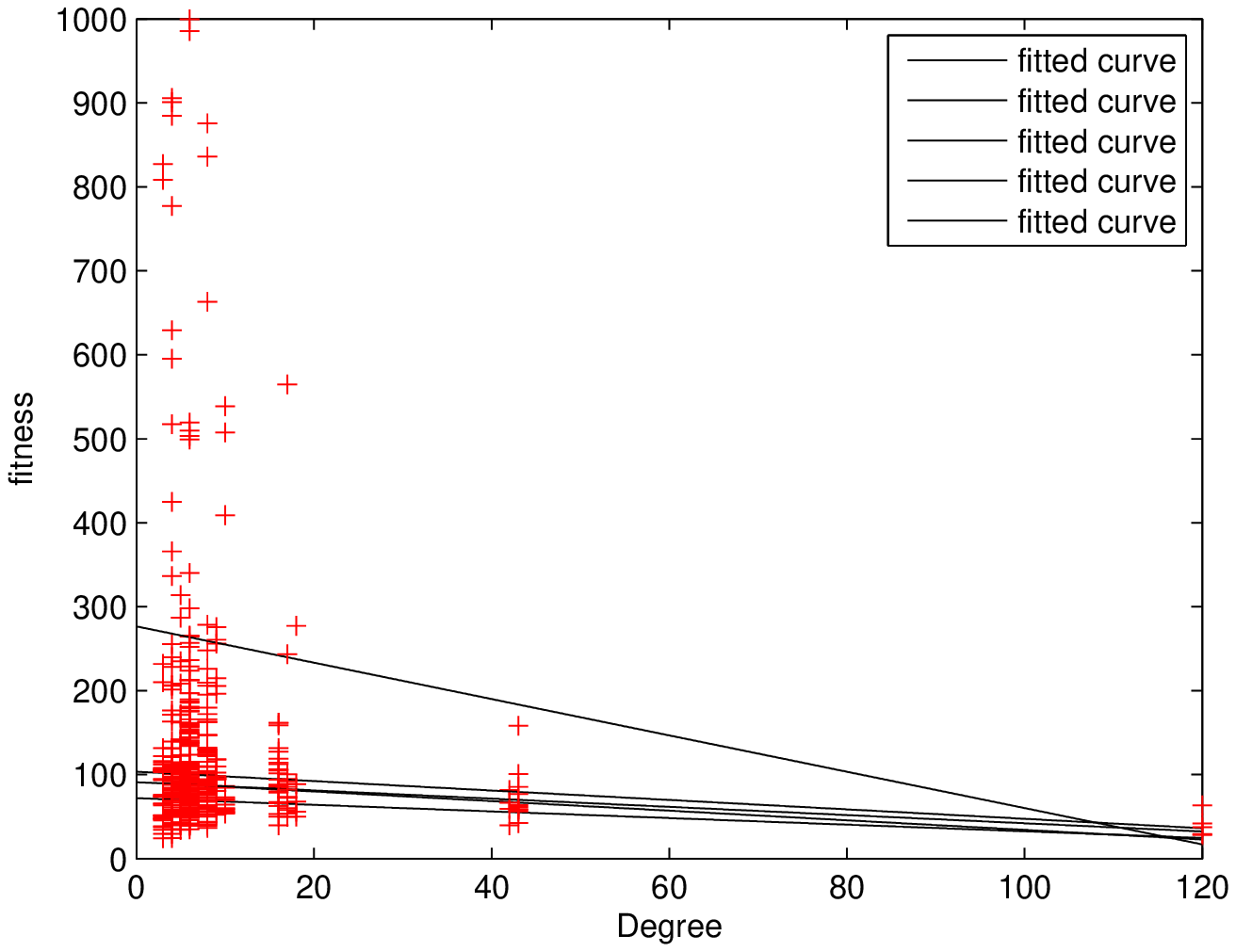}
\includegraphics[scale=0.3]{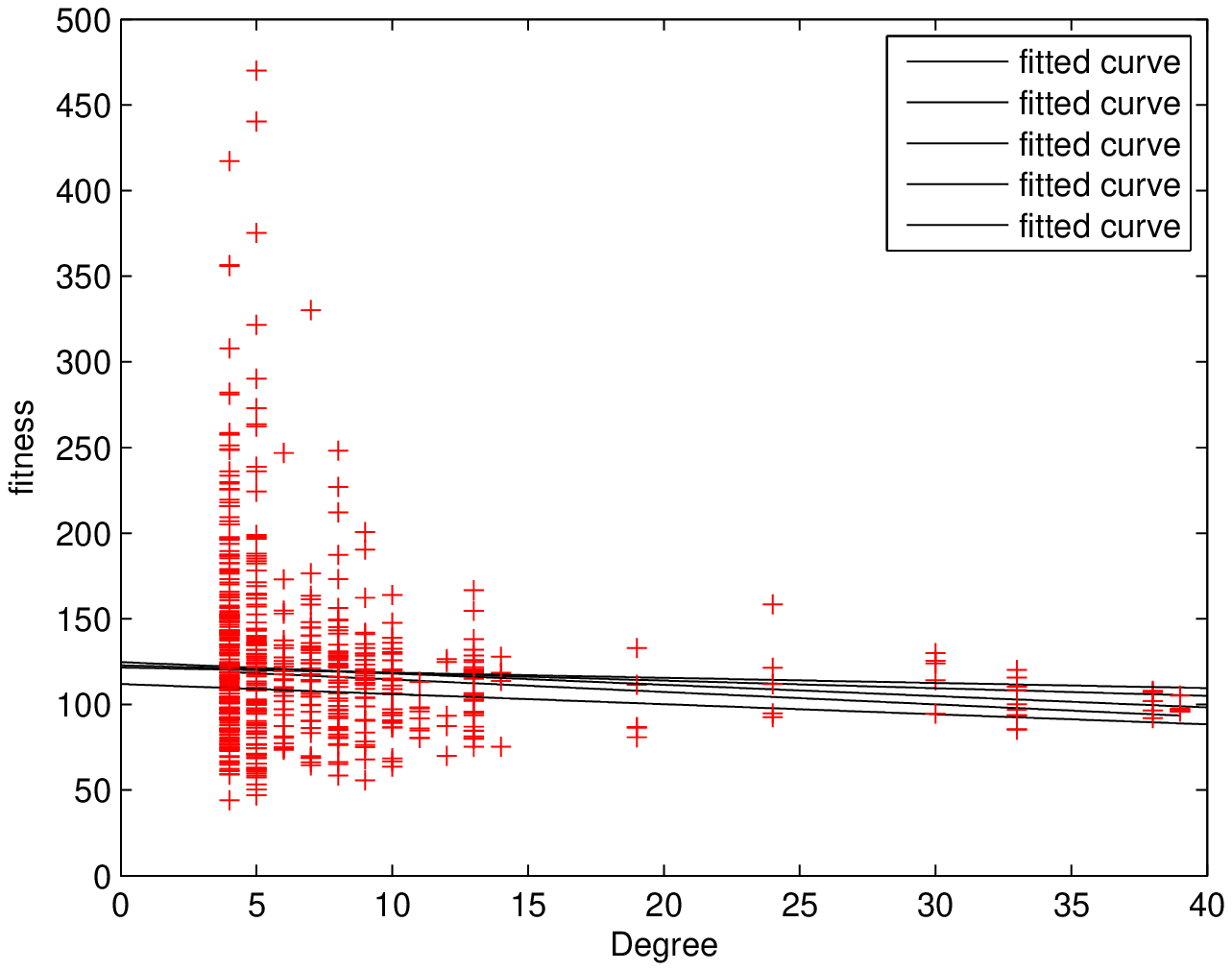}
\includegraphics[scale=0.3]{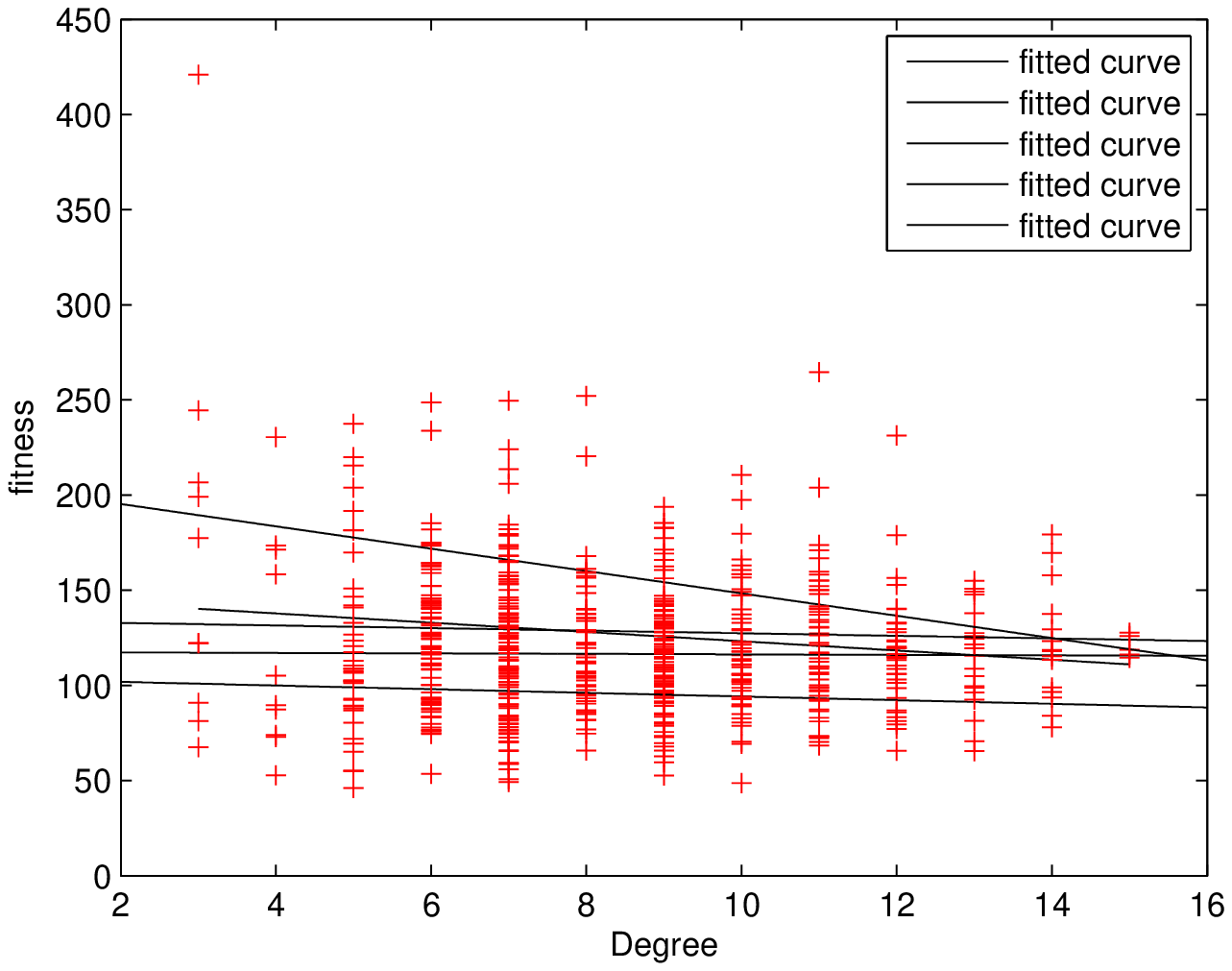}
\caption{Fitness against degree in resp hierarchical, non-hierarchical and random network.
Slope and $95 \%$ confidence interval:$-0.9795$ and  $[-1.779, -0.1799]$ for hierarchical; $-0.1345$ and $[-1.141, 0.8718]$ for non-hierarchical; $-1.206$ and $[-3.608, 1.196]$ for random network.}\label{fitfriction}
\end{center}
\end{figure}

\begin{figure}[htbf]
\begin{center}
\includegraphics[scale=0.3]{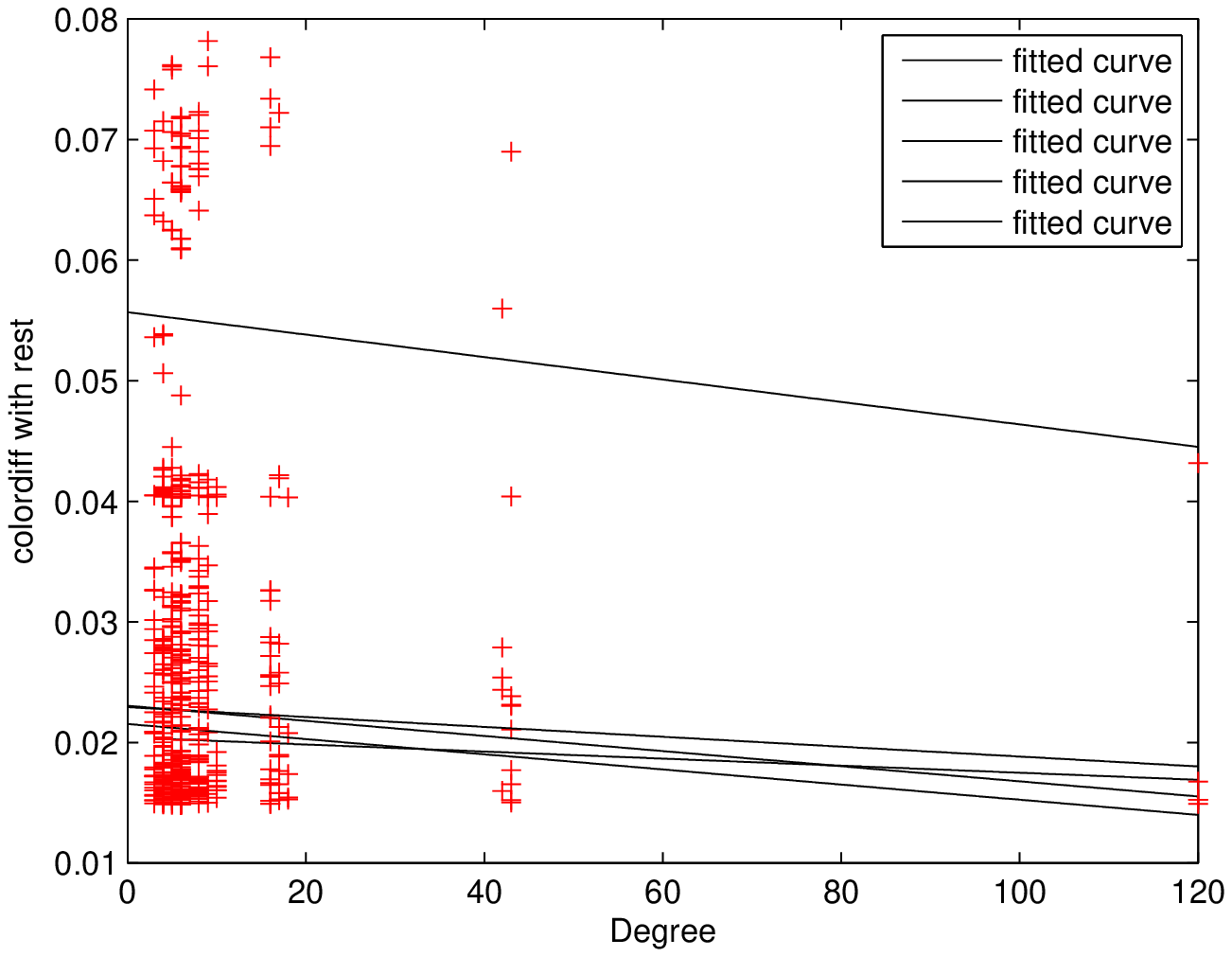}
\includegraphics[scale=0.3]{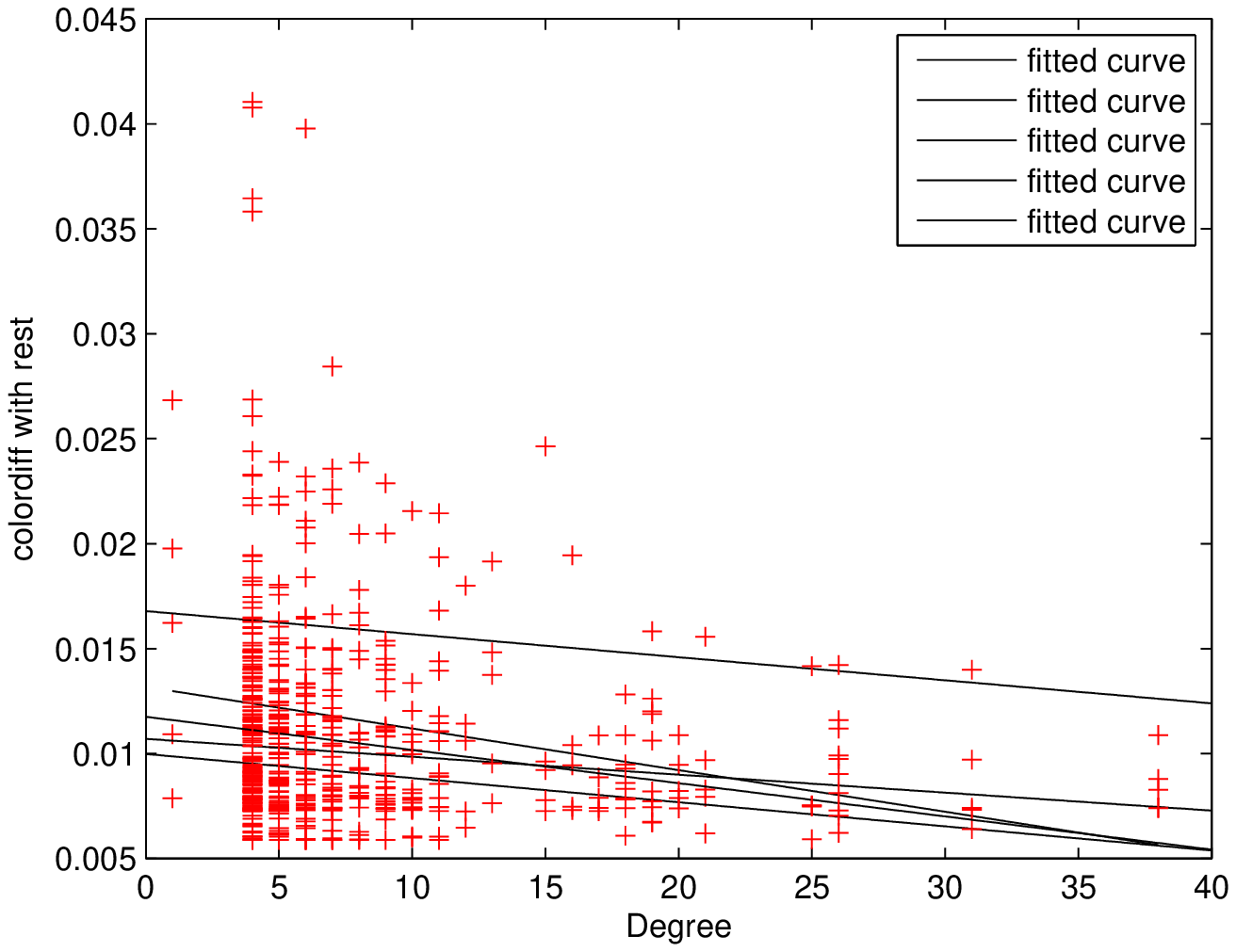}
\includegraphics[scale=0.3]{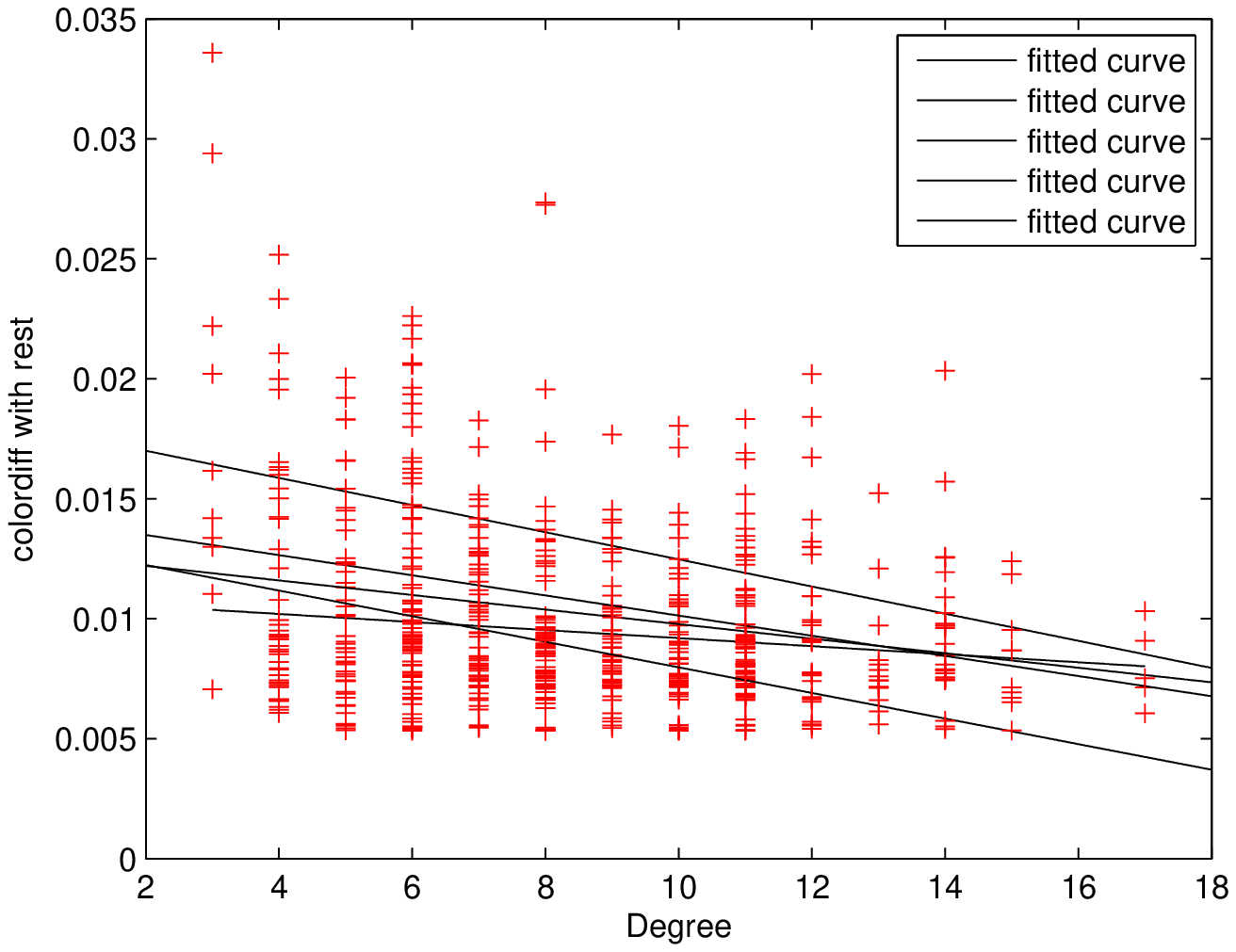}
\caption{Colordifference against degree in resp hierarchical, non-hierarchical and random network. Slope and $95 \%$ confidence interval: $-3.552e-005$ and $[-5.986e-005, -1.119e-005]$ for hierarchical;  $-0.0001232$ and $[-0.0002117, -3.466e-005]$ for non-hierarchical; $-0.0002168$  and $[-0.0003554, -7.822e-005]$ for random network.}\label{colorfriction}
\end{center}
\end{figure}

\newpage

\begin{figure}[htbf]
\begin{center}
\includegraphics[scale=0.4]{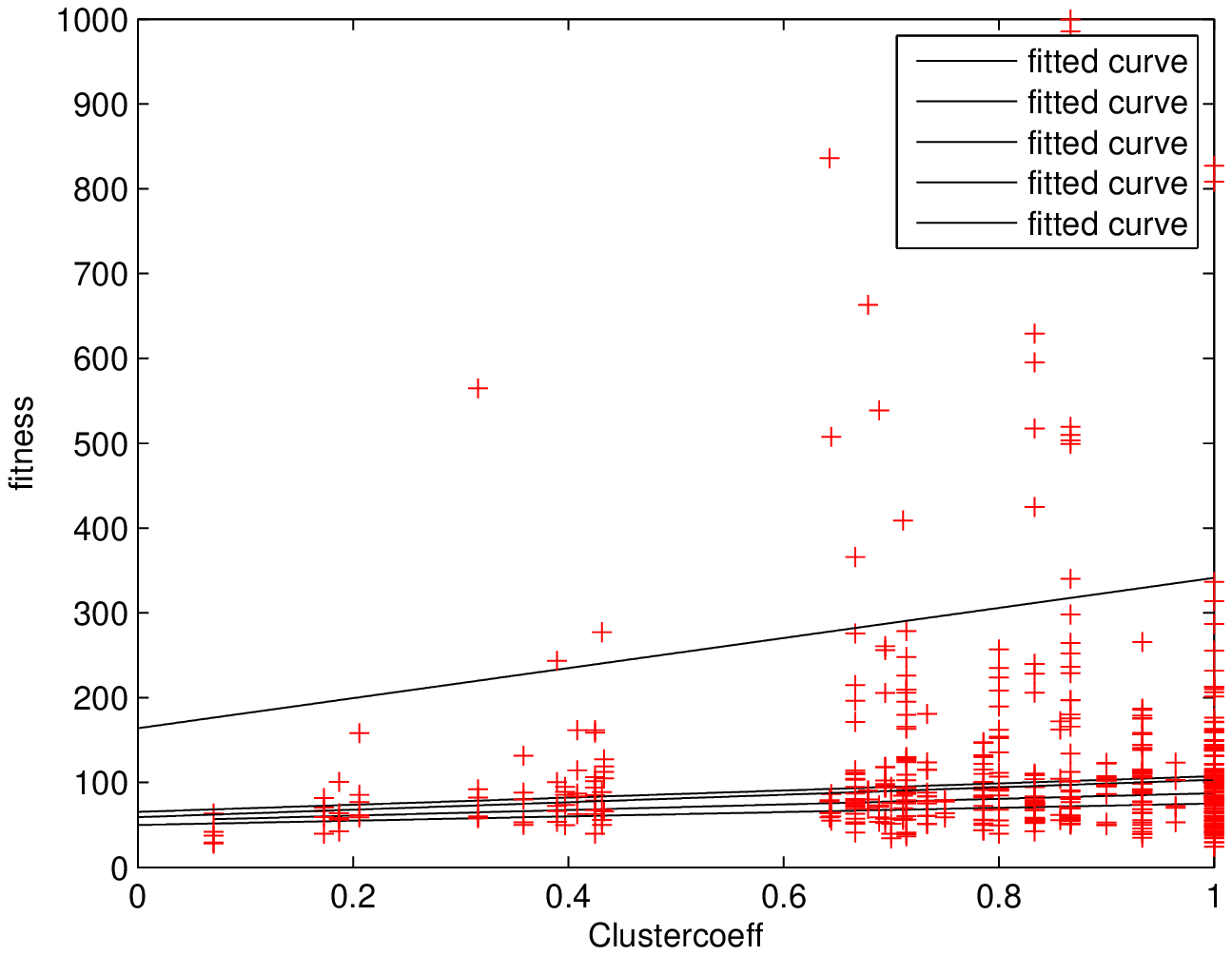}
\includegraphics[scale=0.4]{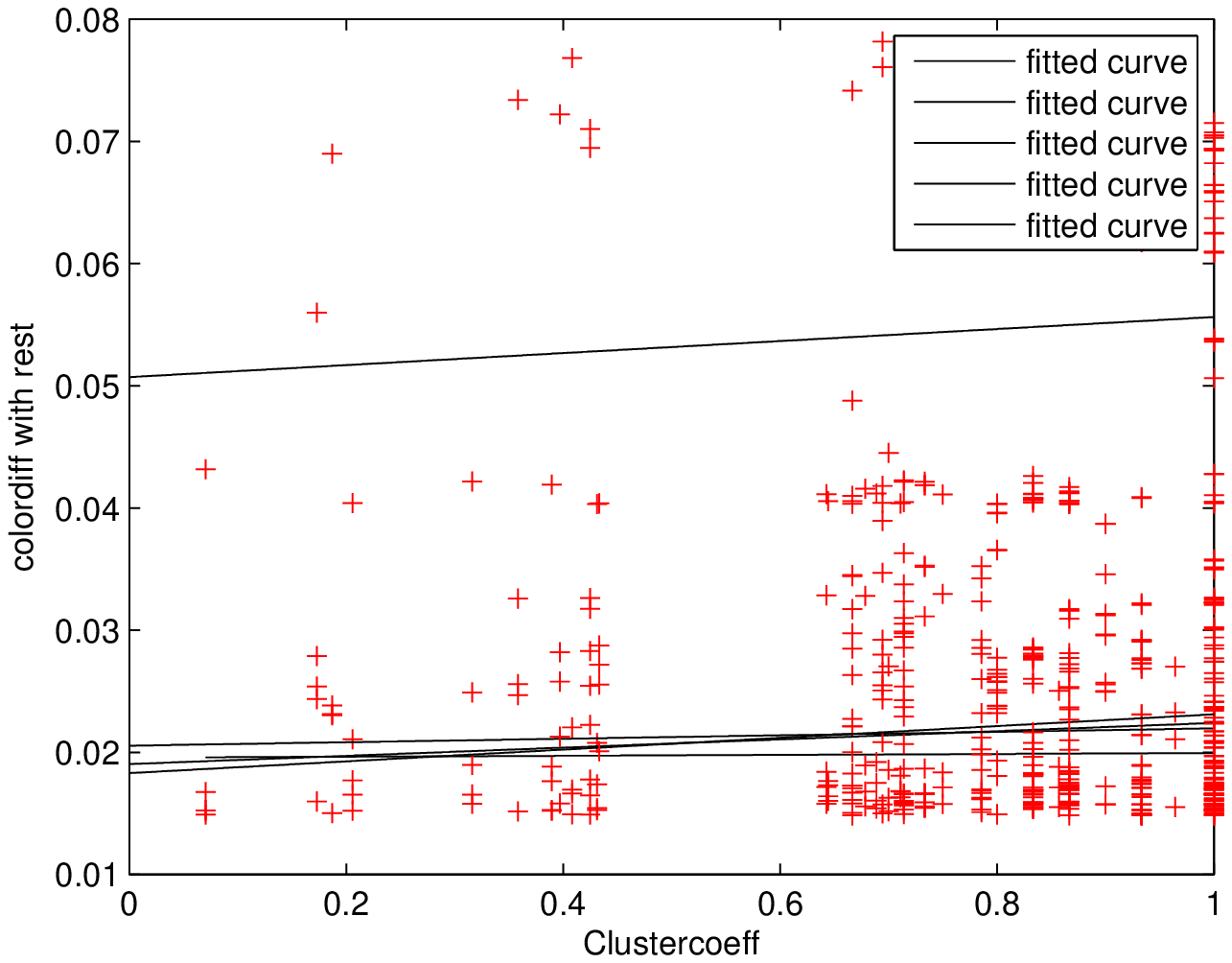}
\caption{Fitness and Colordifference against cluster coefficient in hierarchical network. Slope and $95 \%$ confidence interval: $100.5$ and $[39.73, 161.2]$ against fitness; $0.003266$ and $[-0.005879, 0.01241]$ against colordiff.}\label{cchfriction}
\end{center}
\end{figure}

\begin{figure}[htbf]
\begin{center}
\includegraphics[scale=0.4]{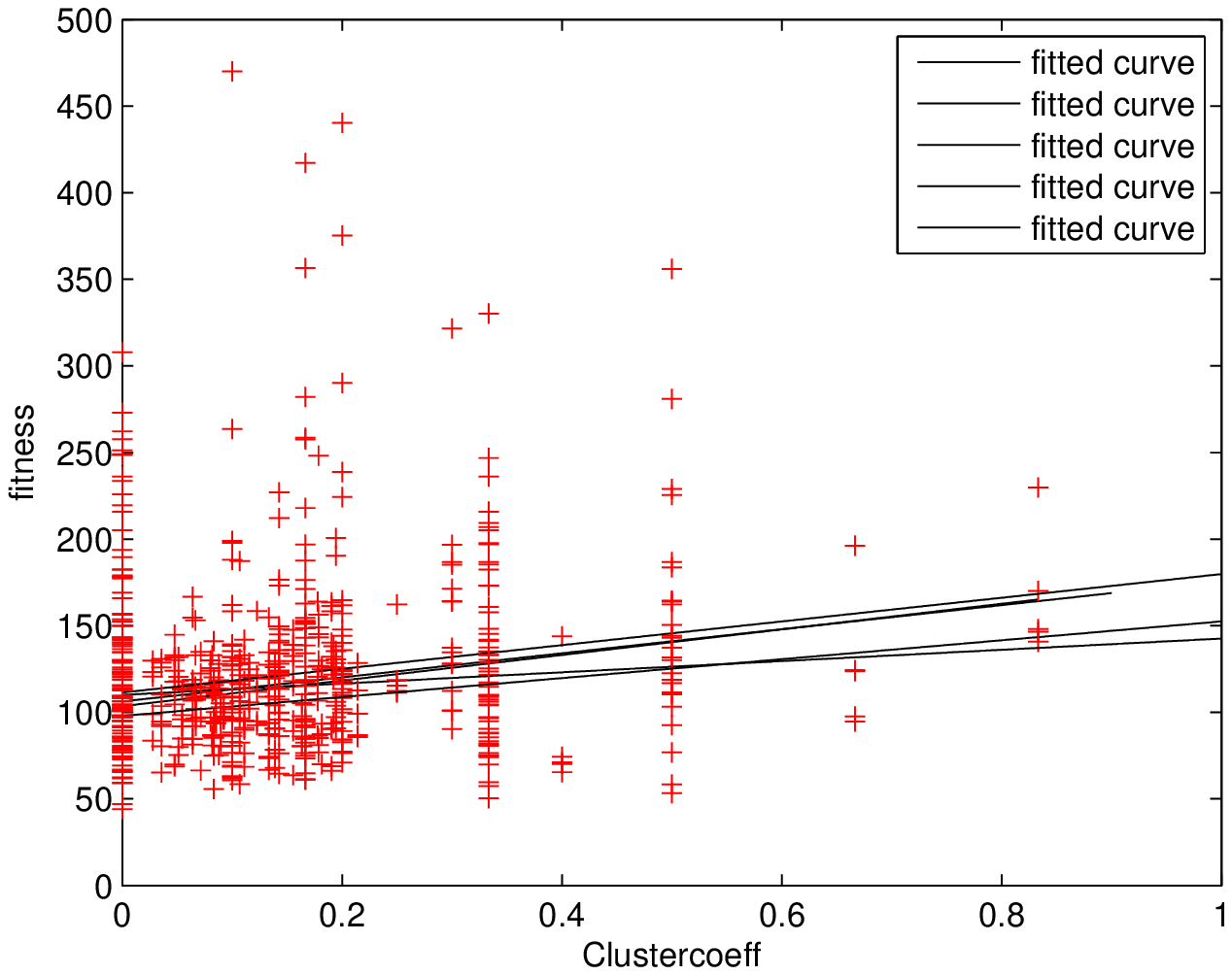}
\includegraphics[scale=0.4]{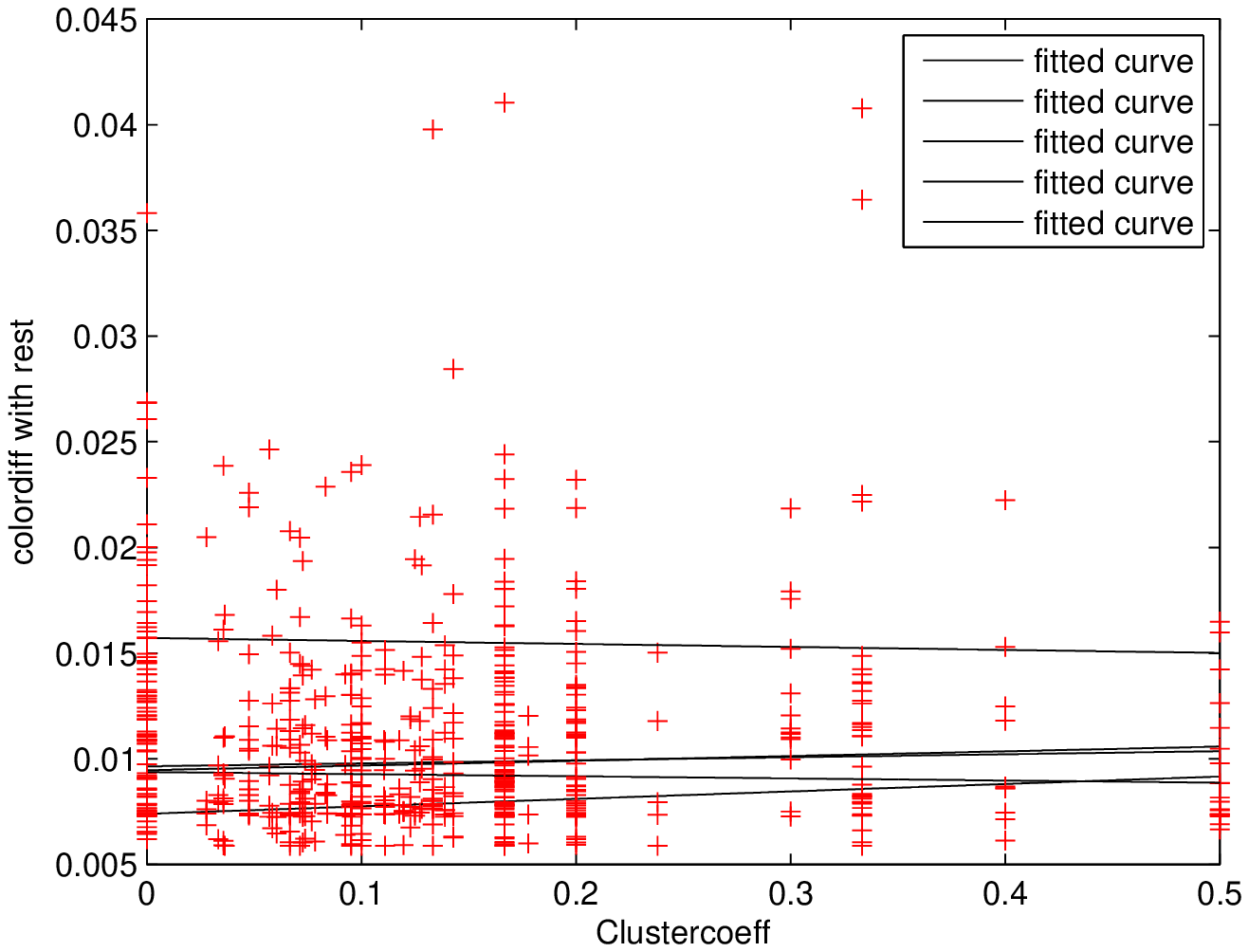}
\caption{Fitness and Colordifference against cluster coefficient in non-hierarchical network. Slope and $95 \%$ confidence interval: $134.6$ and$[71.82, 197.3]$ against fitness; $-0.0006875$ and $[-0.0123, 0.01092]$ against colordiff.}\label{ccnhfriction}
\end{center}
\end{figure}

\begin{figure}[htbf]
\begin{center}
\includegraphics[scale=0.4]{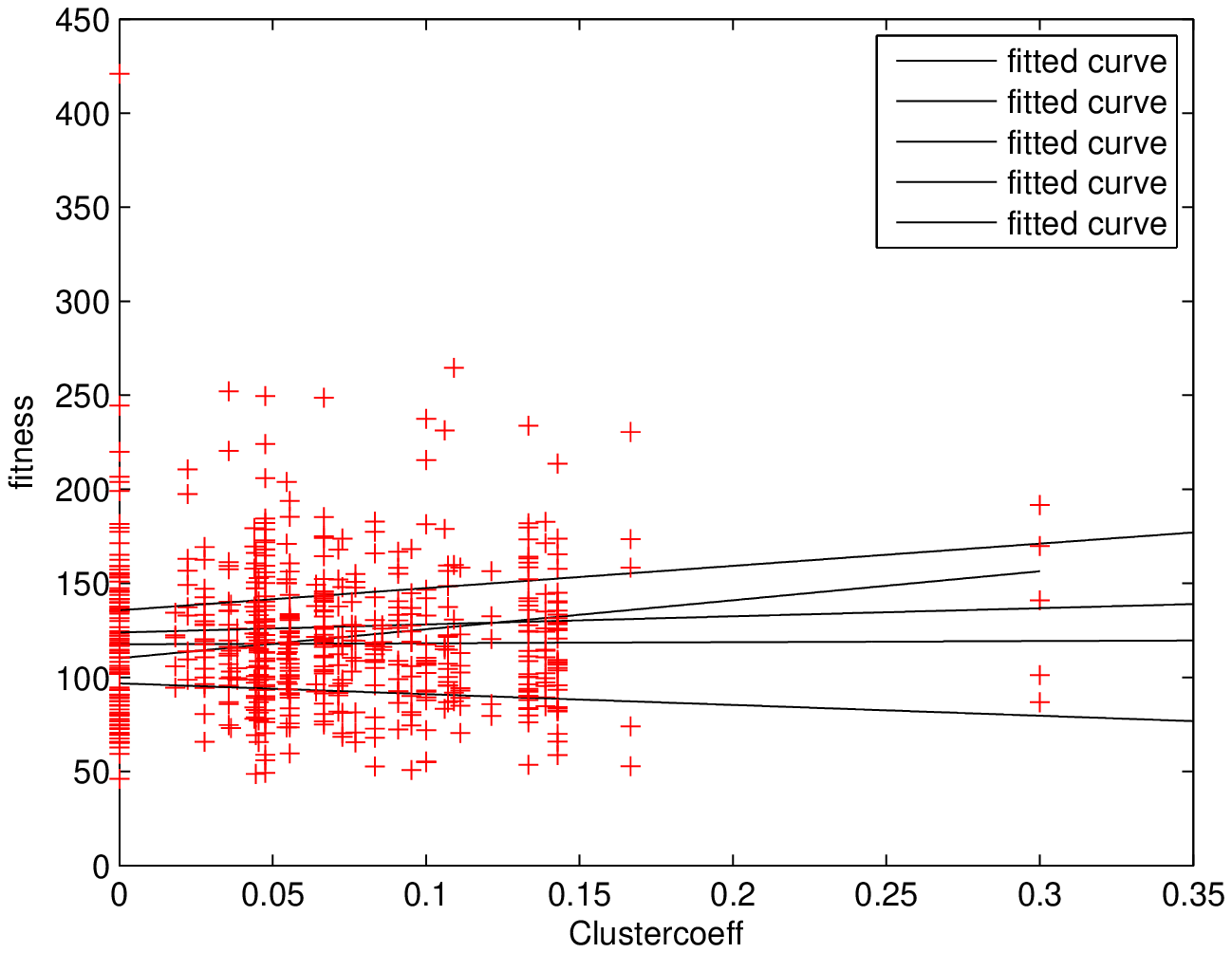}
\includegraphics[scale=0.4]{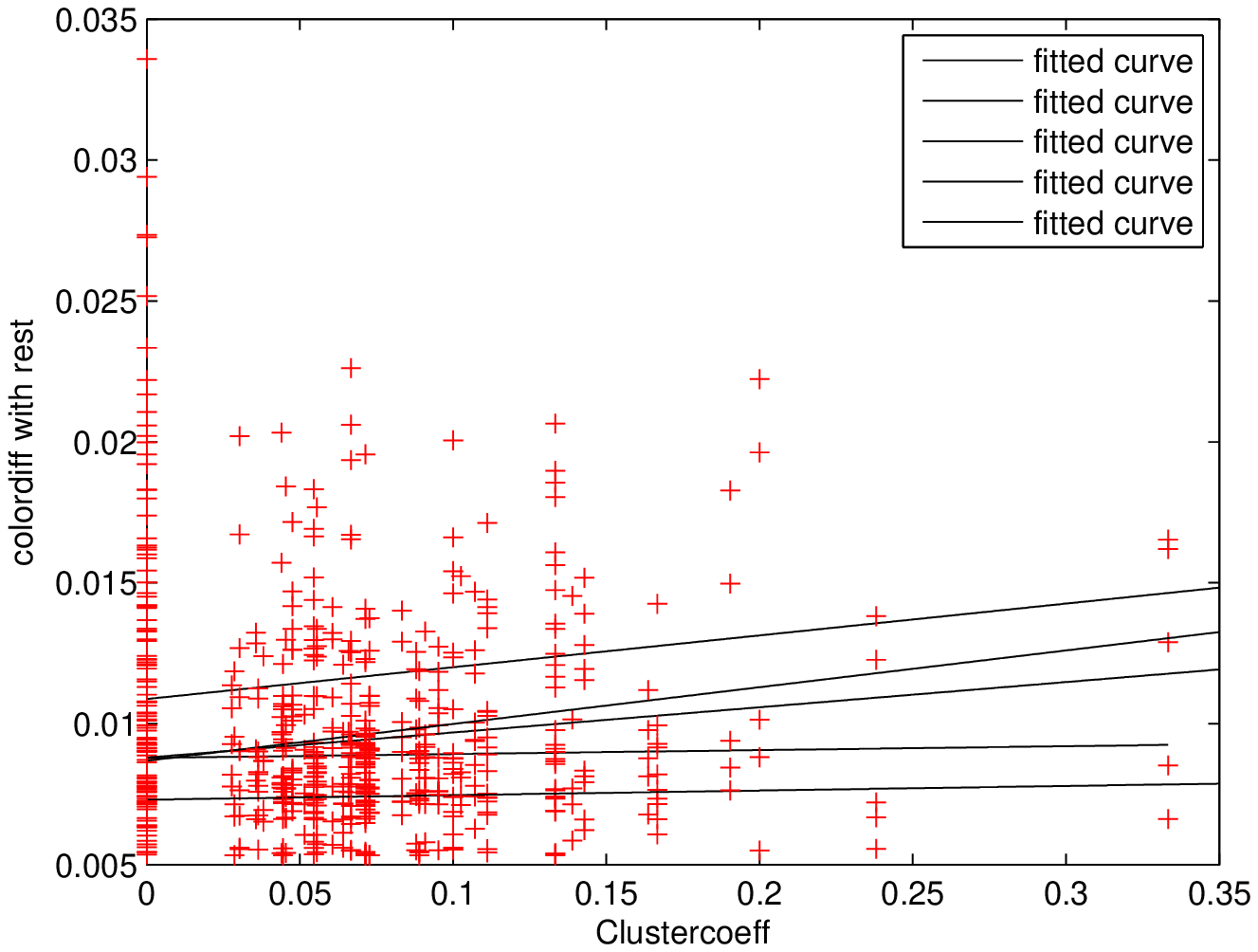}
\caption{Fitness and Colordifference against cluster coefficient in random network.  Slope and $95 \%$ confidence interval: $-78.63$ and $[-260.7, 103.5 ]$ against fitness; $0.01359$ and $[-0.002423, 0.02959]$ against colordiff.}\label{ccrfriction}
\end{center}
\end{figure}

\newpage
\subsection{Fitness in time}\label{fitt}

\begin{figure}[htbf]
\begin{center}
\includegraphics[scale=0.3]{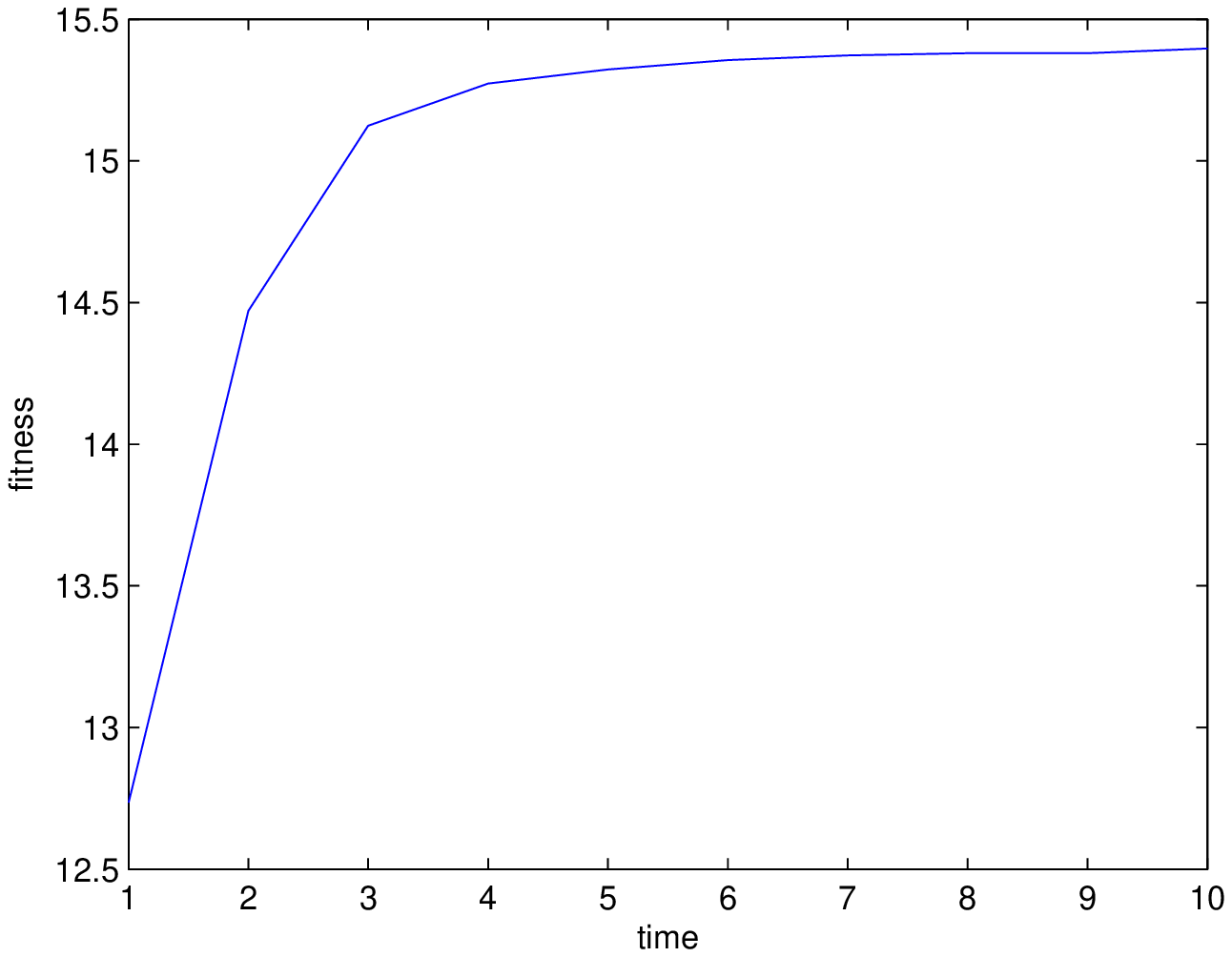}
\includegraphics[scale=0.3]{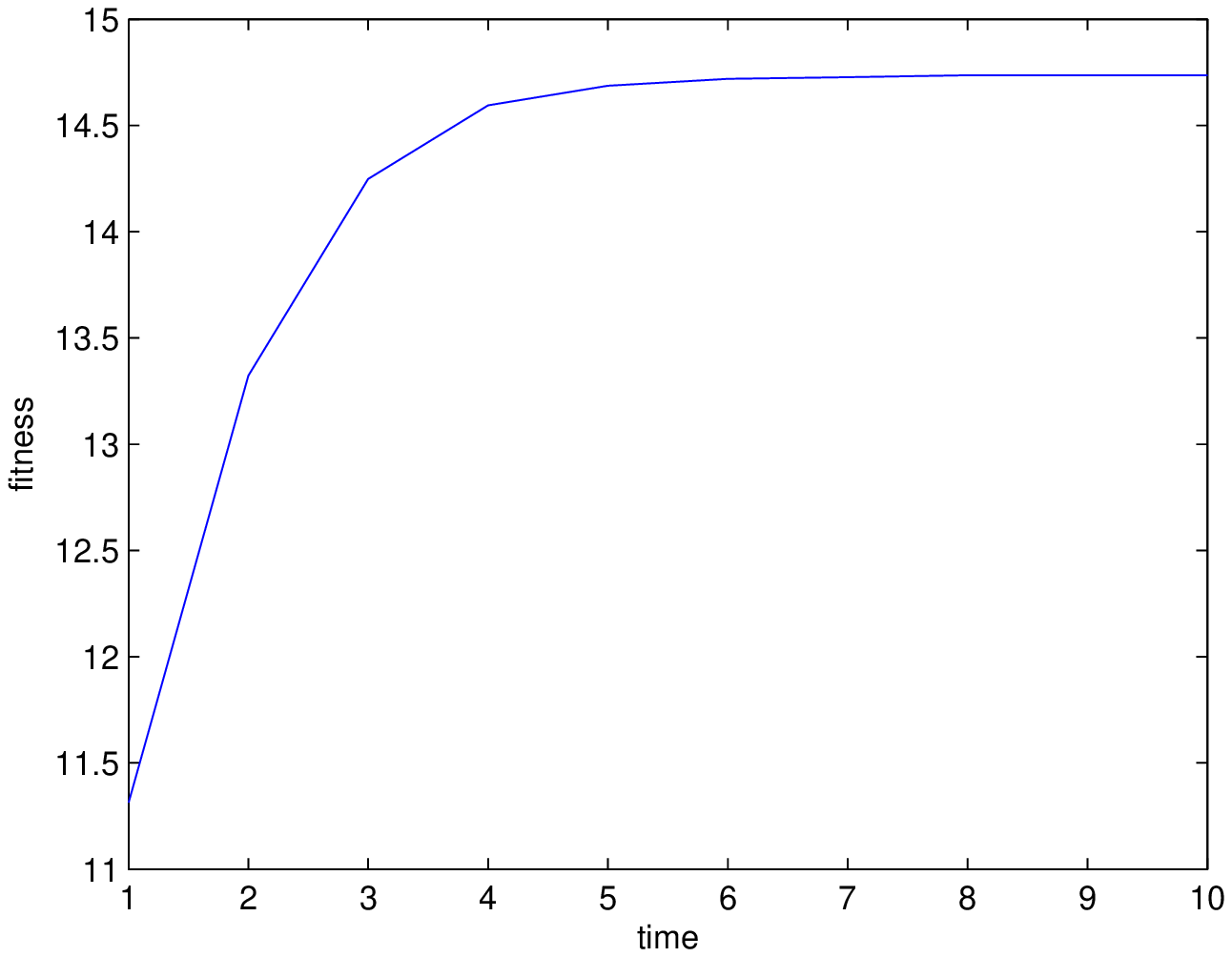}
\includegraphics[scale=0.3]{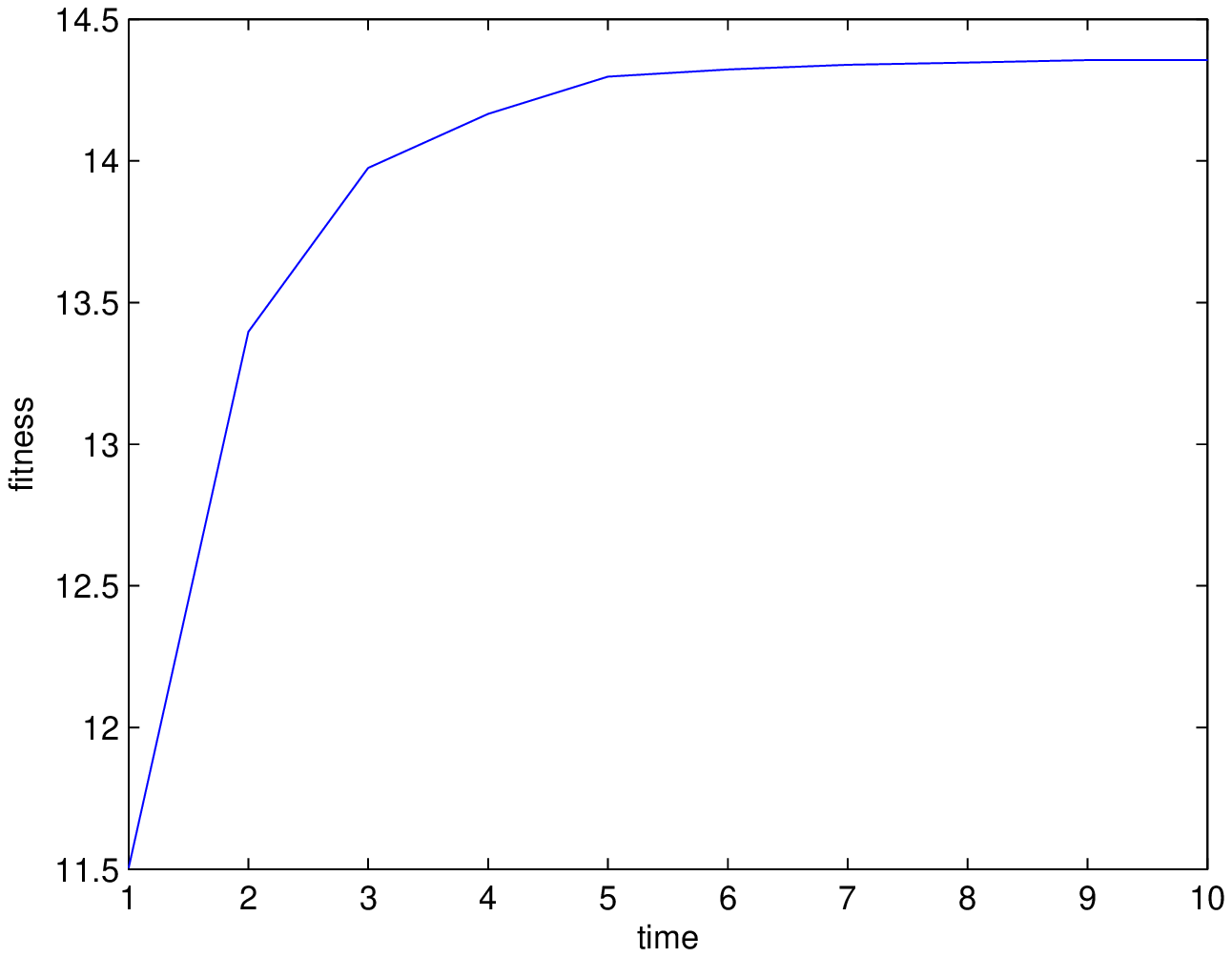}
\caption{Evolution of mean fitness in resp hierarchical, non-hierarchical and random network in non-propagation model}\label{nopropfitt}
\end{center}
\end{figure}

\begin{figure}[htbf]
\begin{center}
\includegraphics[scale=0.3]{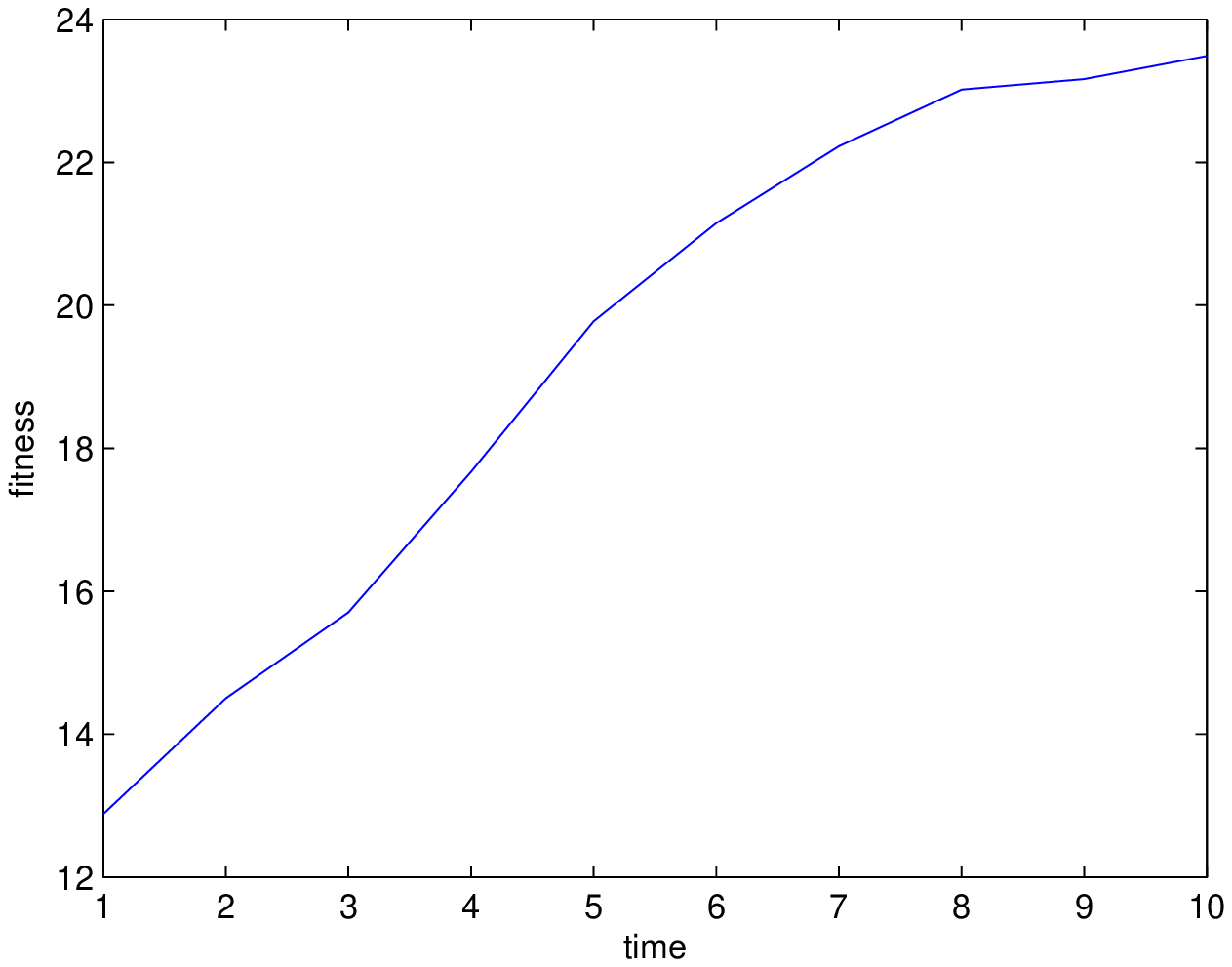}
\includegraphics[scale=0.3]{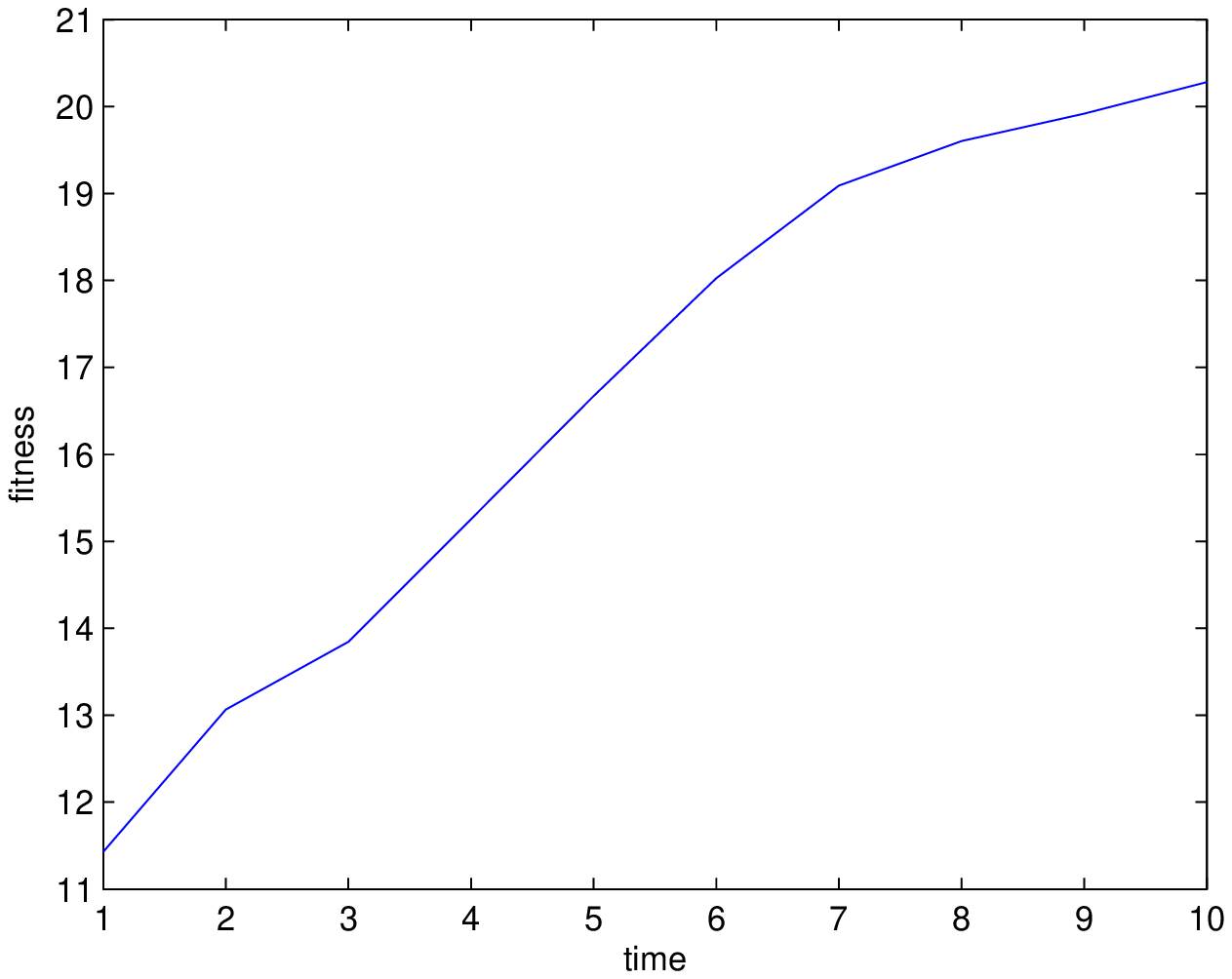}
\includegraphics[scale=0.3]{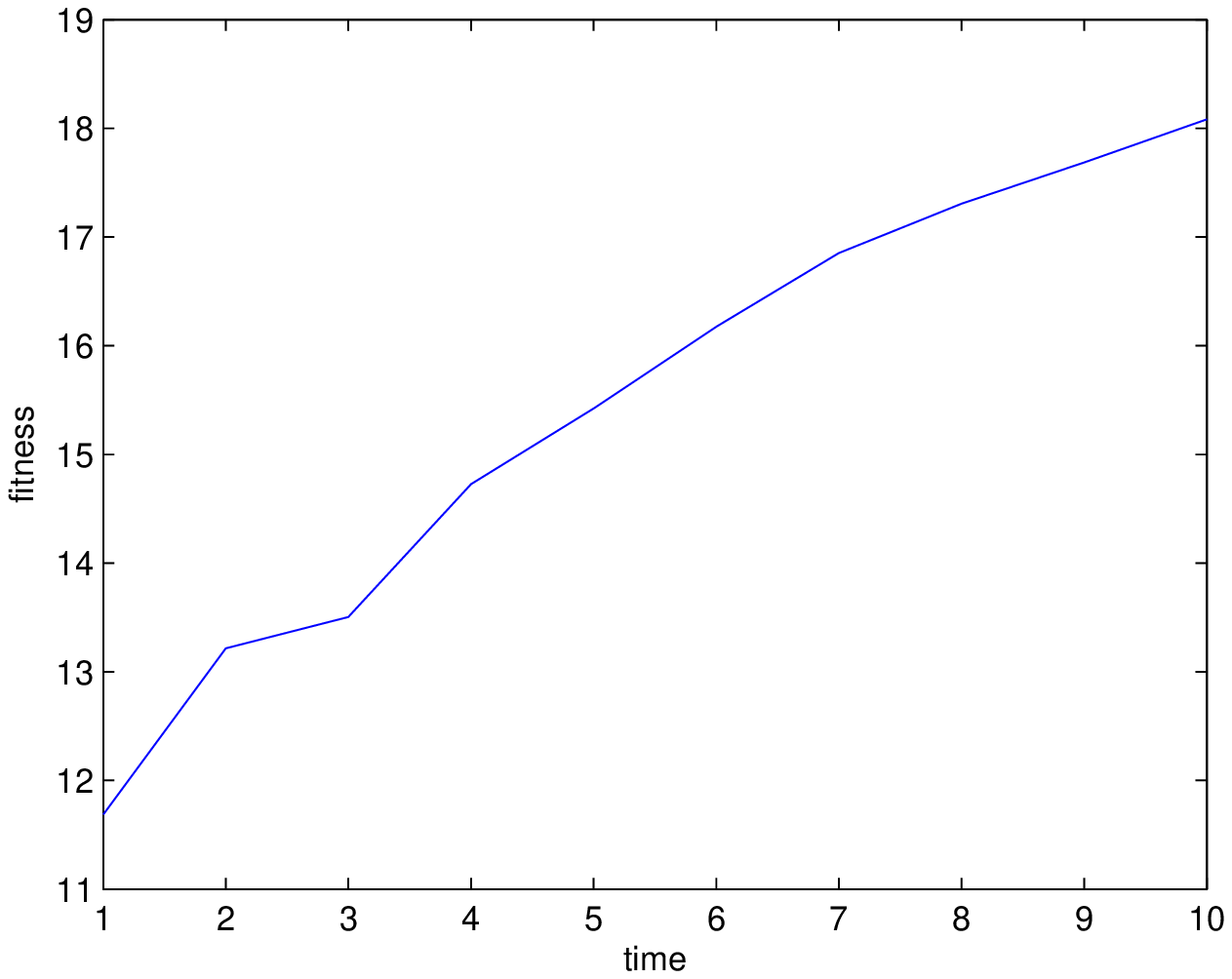}
\caption{Evolution of mean fitness in resp hierarchical, non-hierarchical and random network in propagation model}\label{propfitt}
\end{center}
\end{figure}

\newpage
\subsection{Plots maximize synergy model}\label{synergy}

\subsubsection{with no propagation}
\begin{figure}[htbf]
\begin{center}
\includegraphics[scale=0.3]{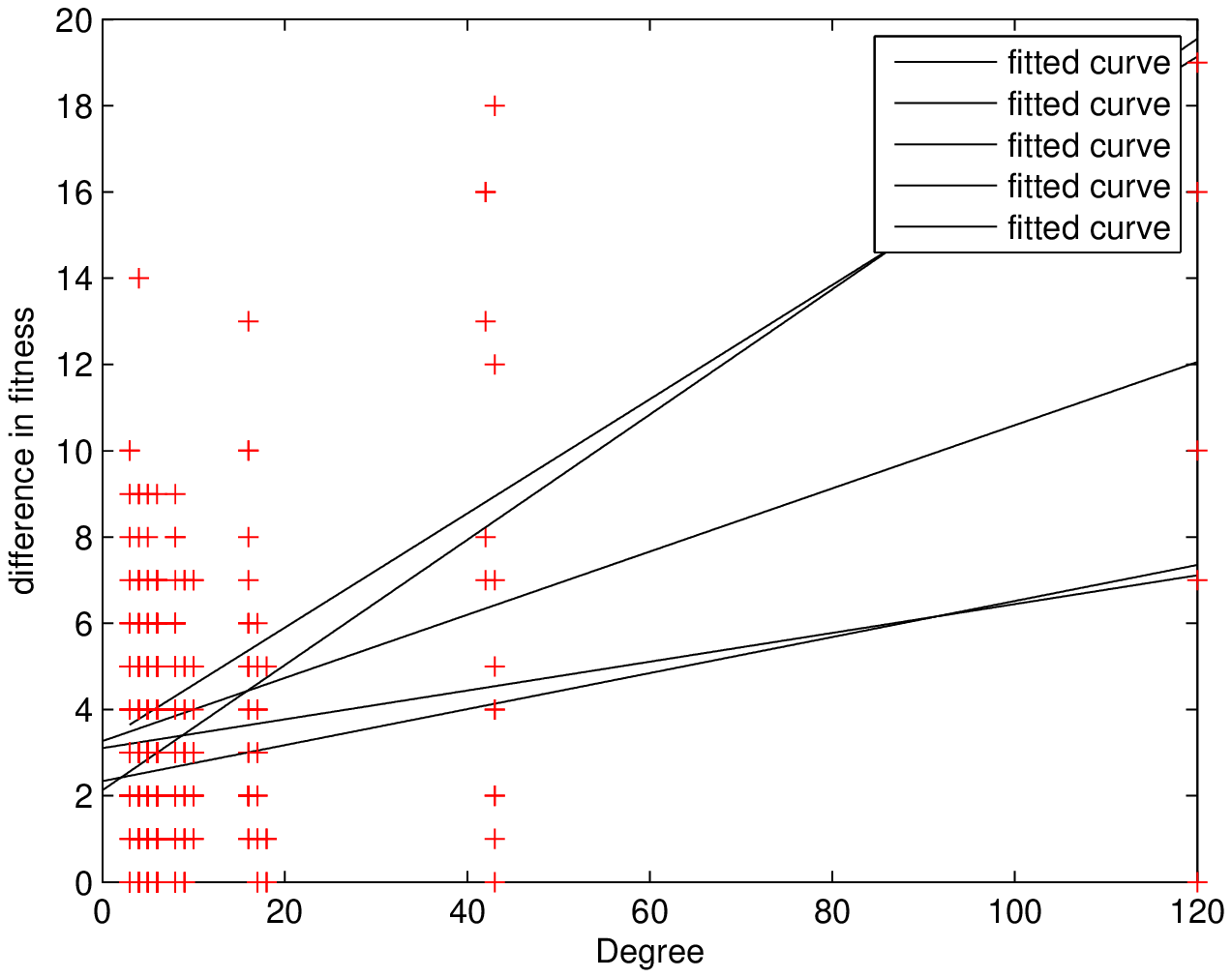}
\includegraphics[scale=0.3]{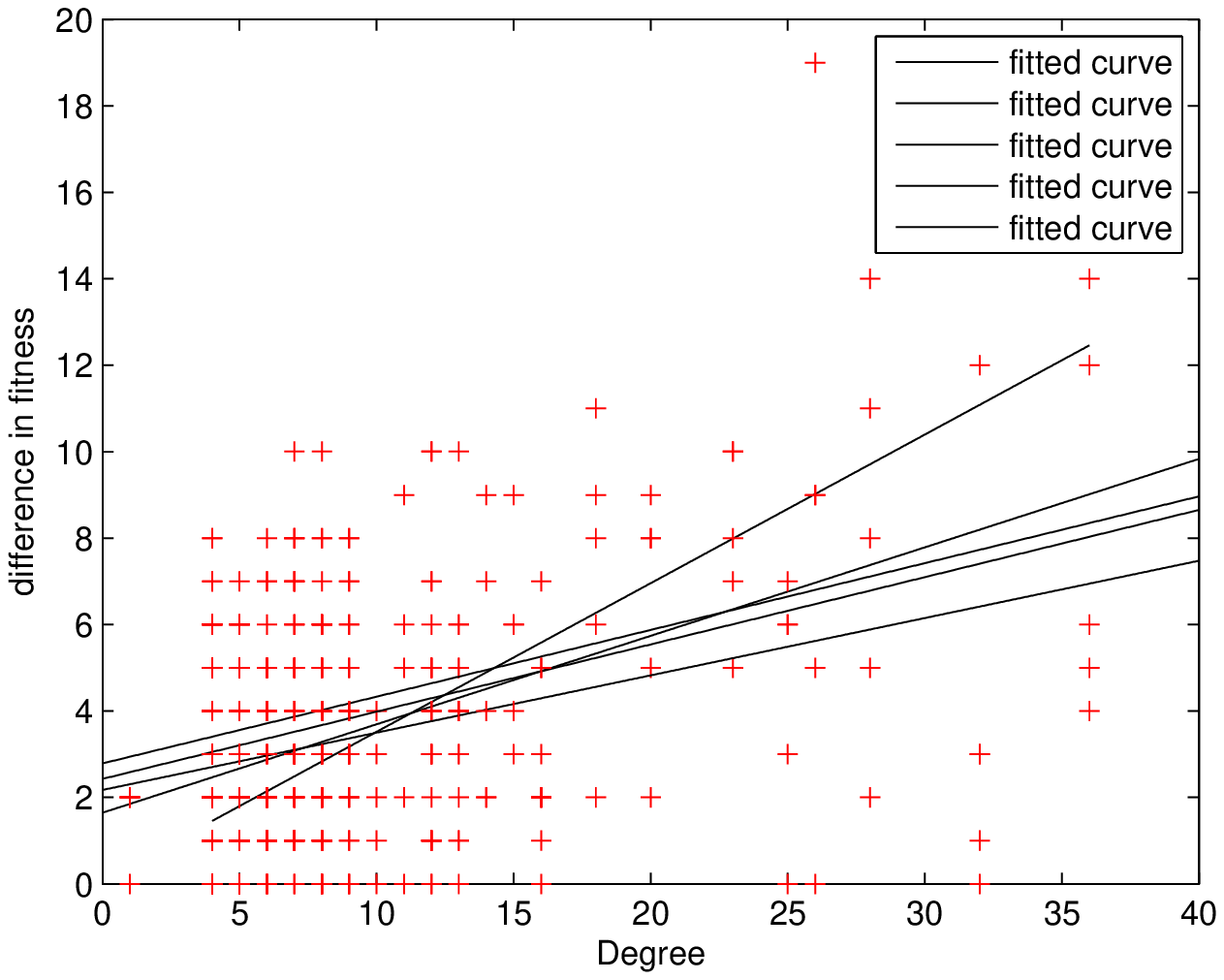}
\includegraphics[scale=0.3]{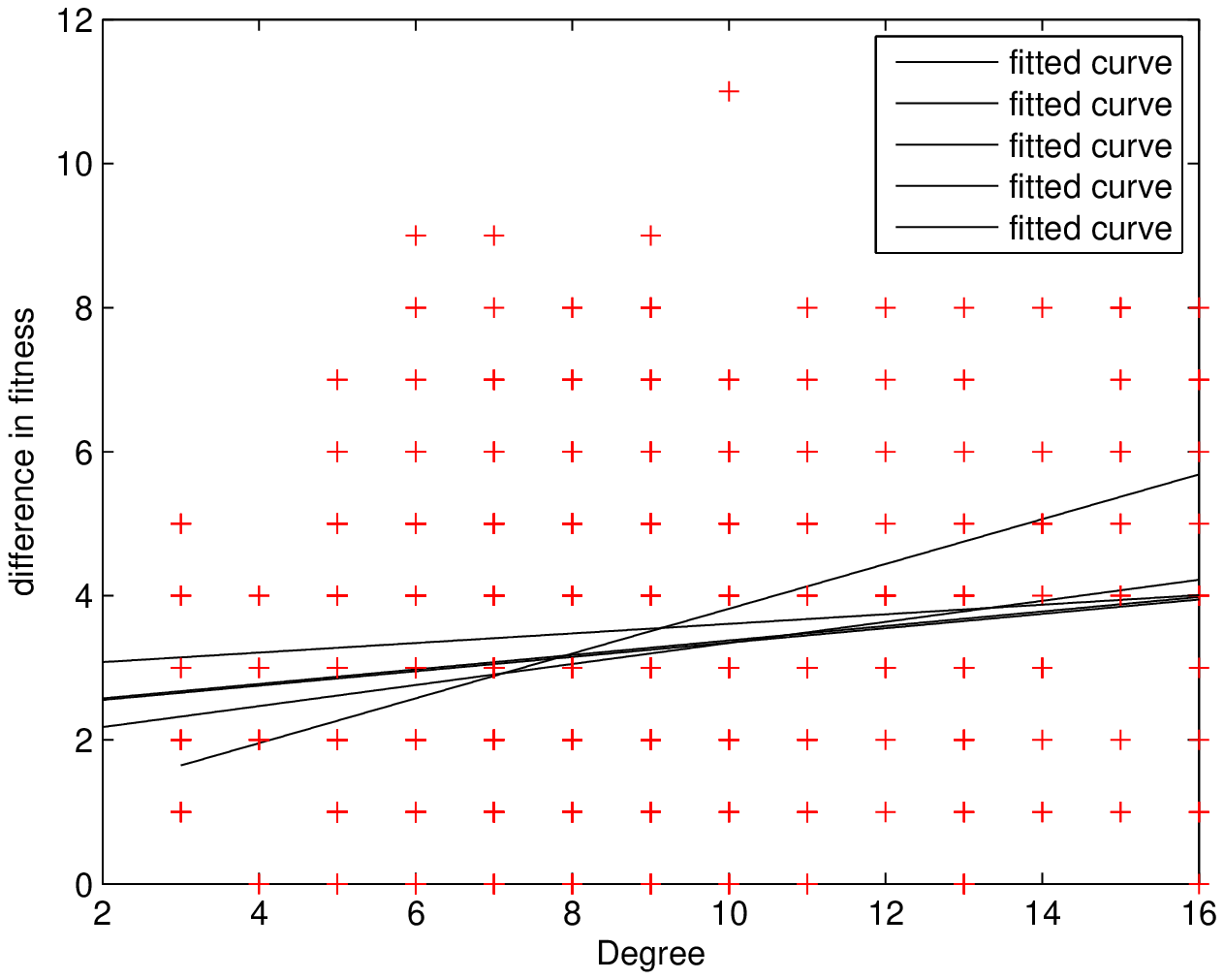}
\caption{Fitness against degree in resp hierarchical, non-hierarchical and random network}
\end{center}
\end{figure}

\begin{figure}[htbf]
\begin{center}
\includegraphics[scale=0.3]{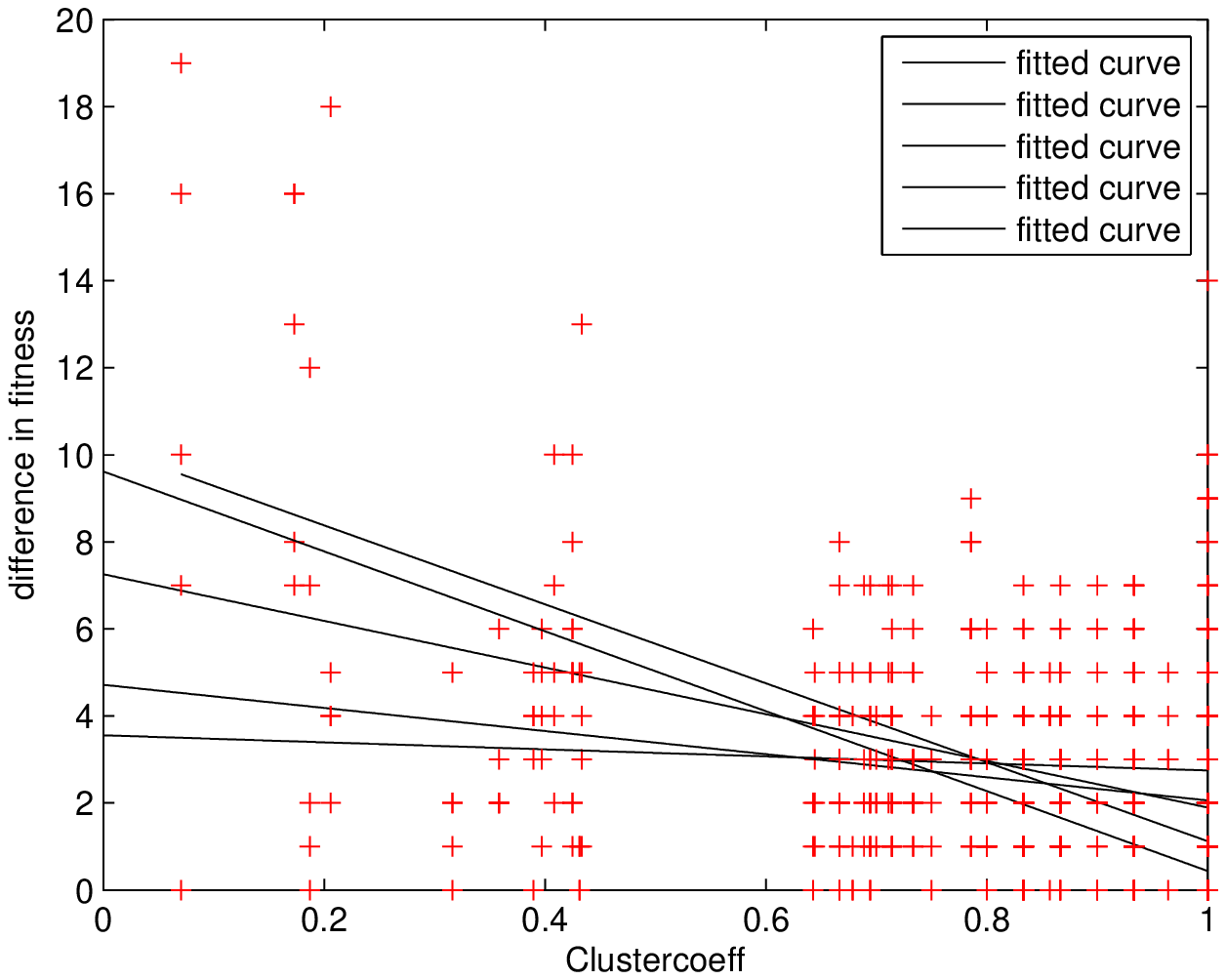}
\includegraphics[scale=0.3]{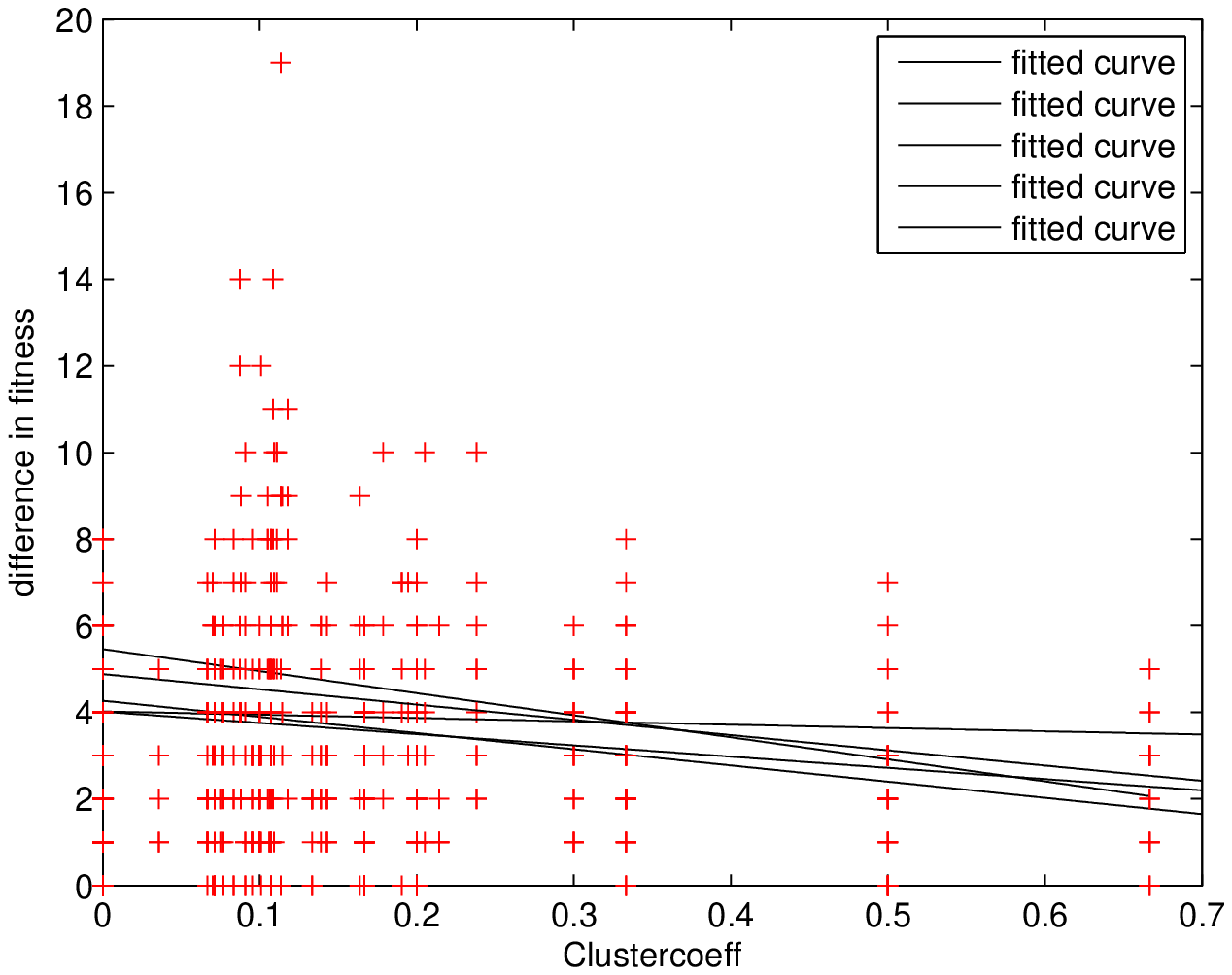}
\includegraphics[scale=0.3]{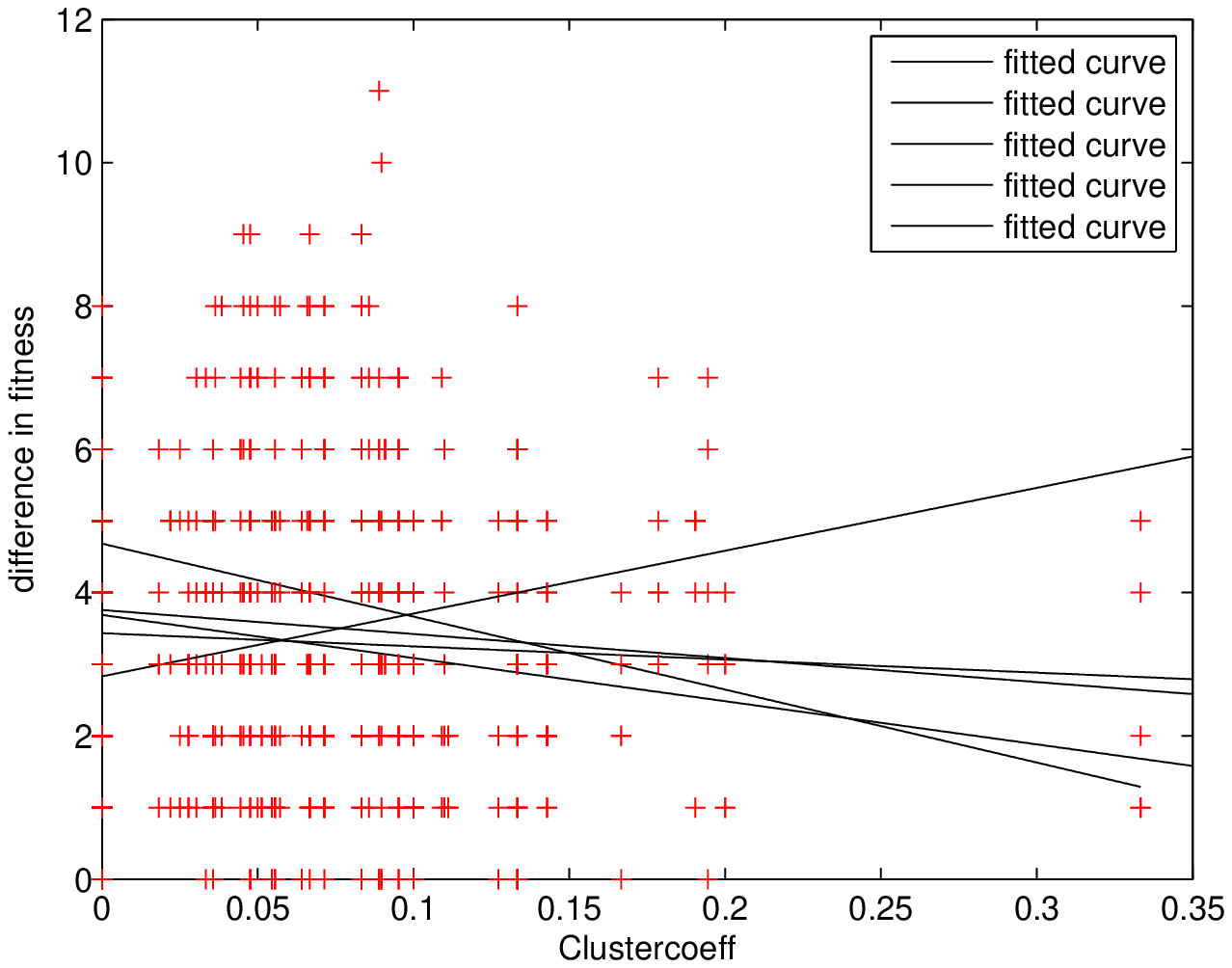}
\caption{Fitness against cluster coefficient in resp hierarchical, non-hierarchical and random network}
\end{center}
\end{figure}

\newpage
\subsubsection{with propagation}
\begin{figure}[htbf]
\begin{center}
\includegraphics[scale=0.3]{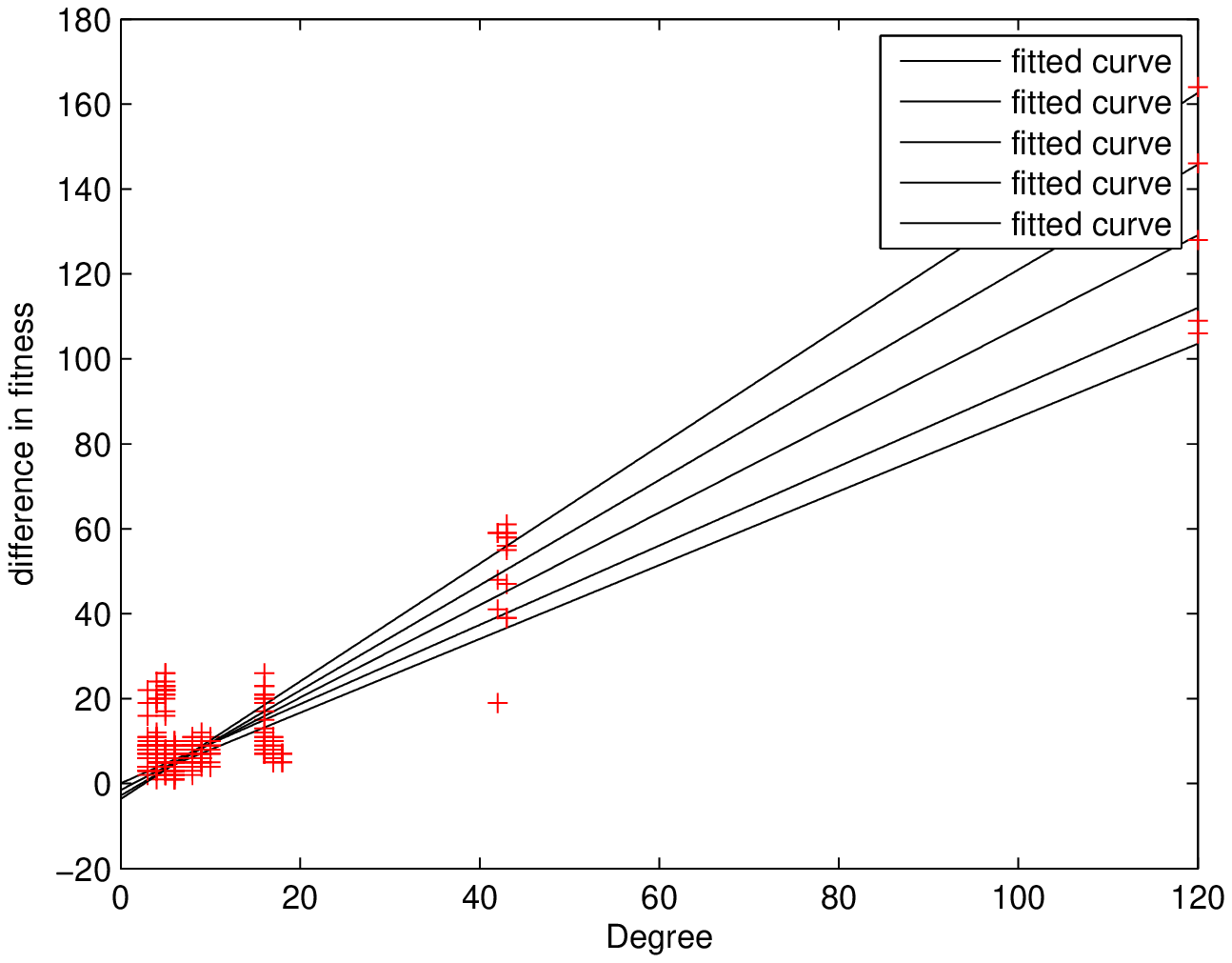}
\includegraphics[scale=0.3]{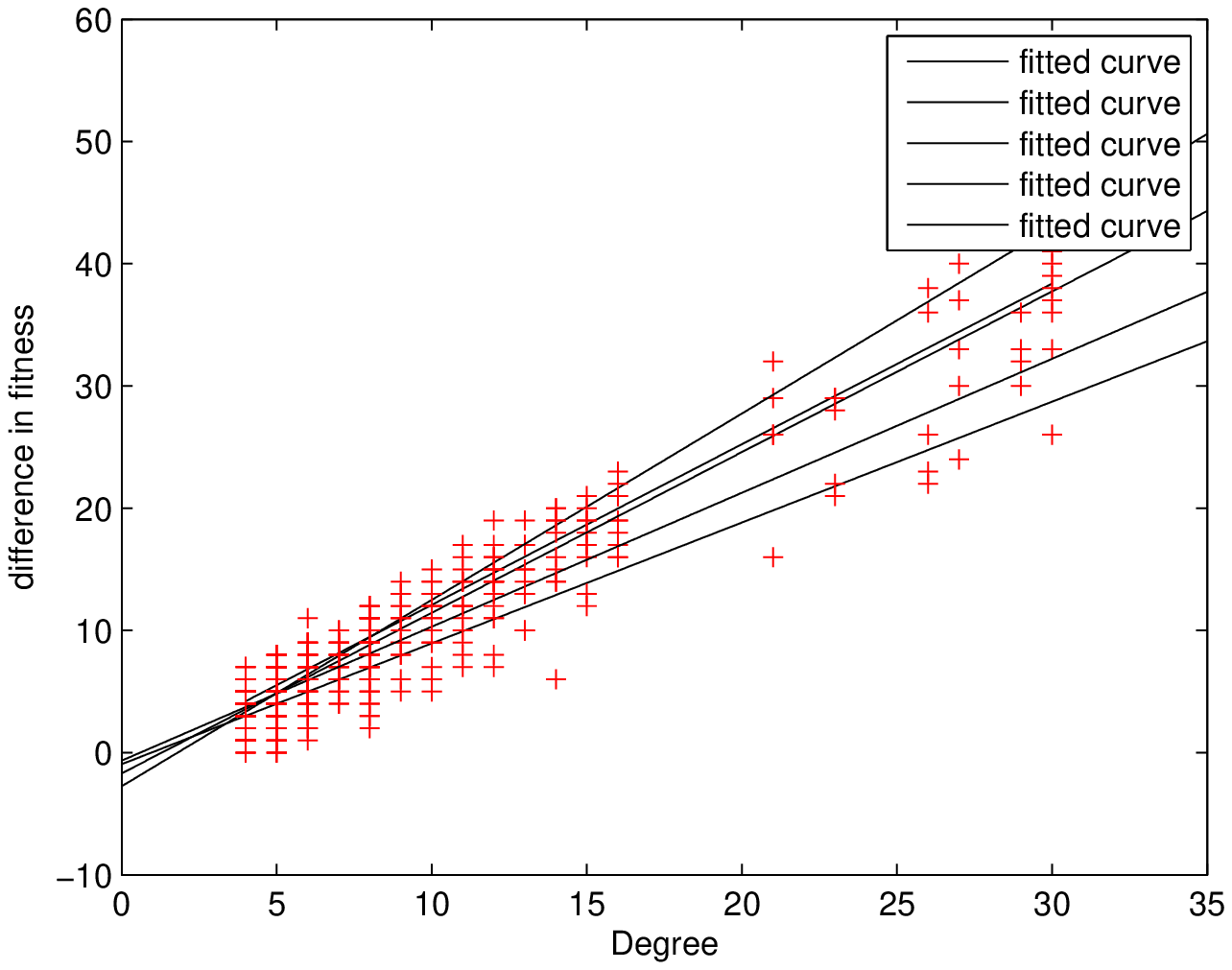}
\includegraphics[scale=0.3]{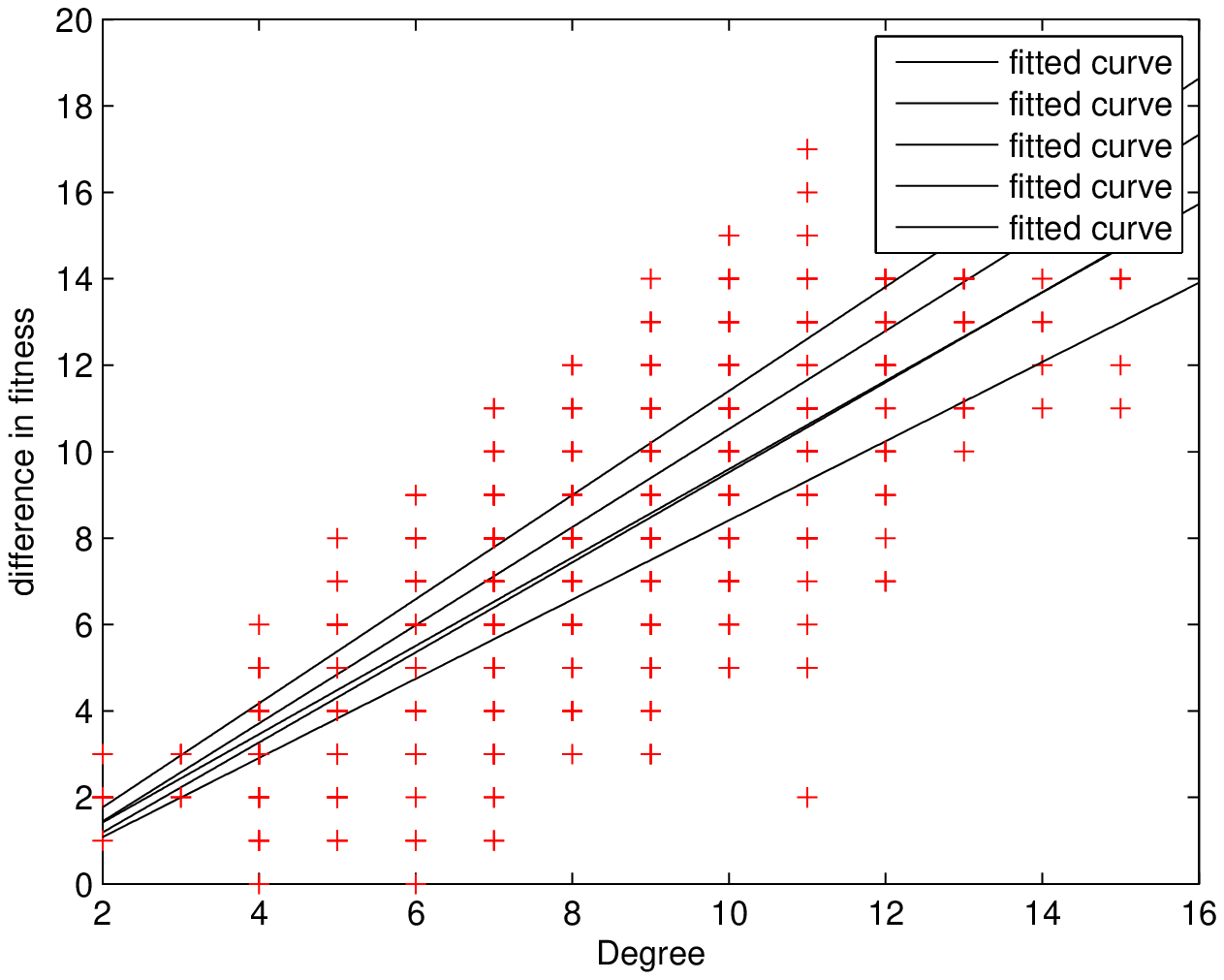}
\caption{Fitness against degree in reps hierarchical, non-hierarchical and random network}
\end{center}
\end{figure}

\begin{figure}[htbf]
\begin{center}
\includegraphics[scale=0.3]{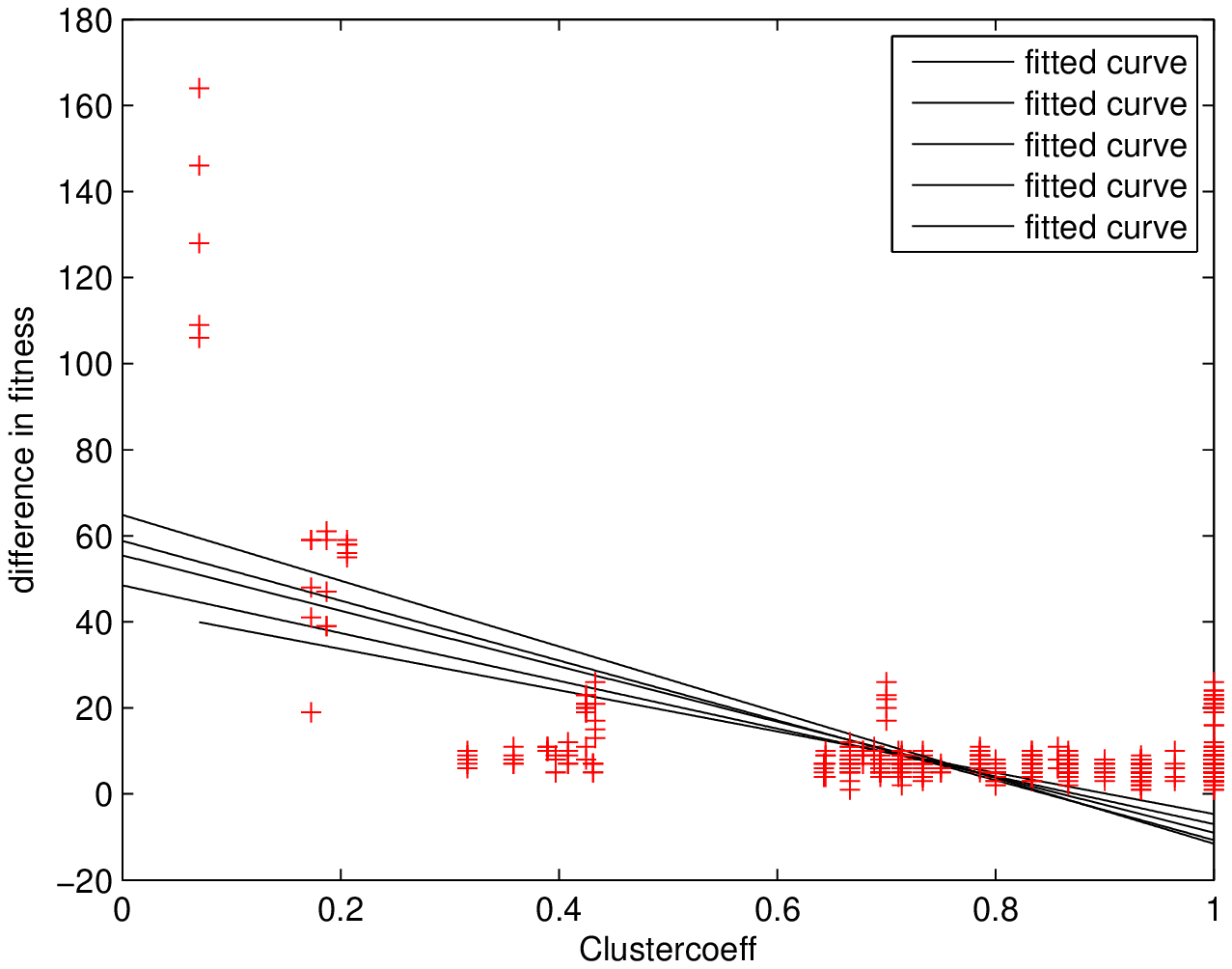}
\includegraphics[scale=0.3]{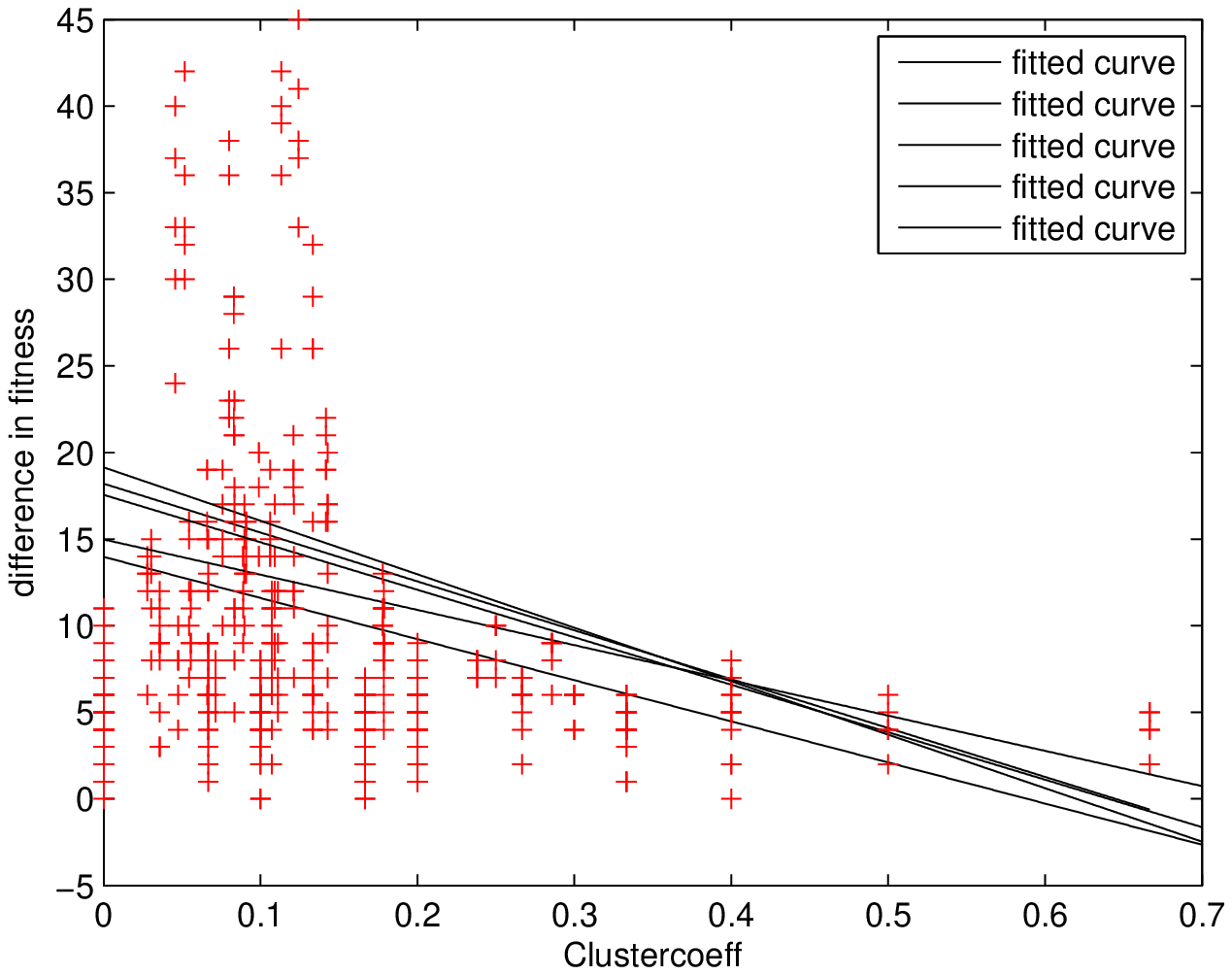}
\includegraphics[scale=0.3]{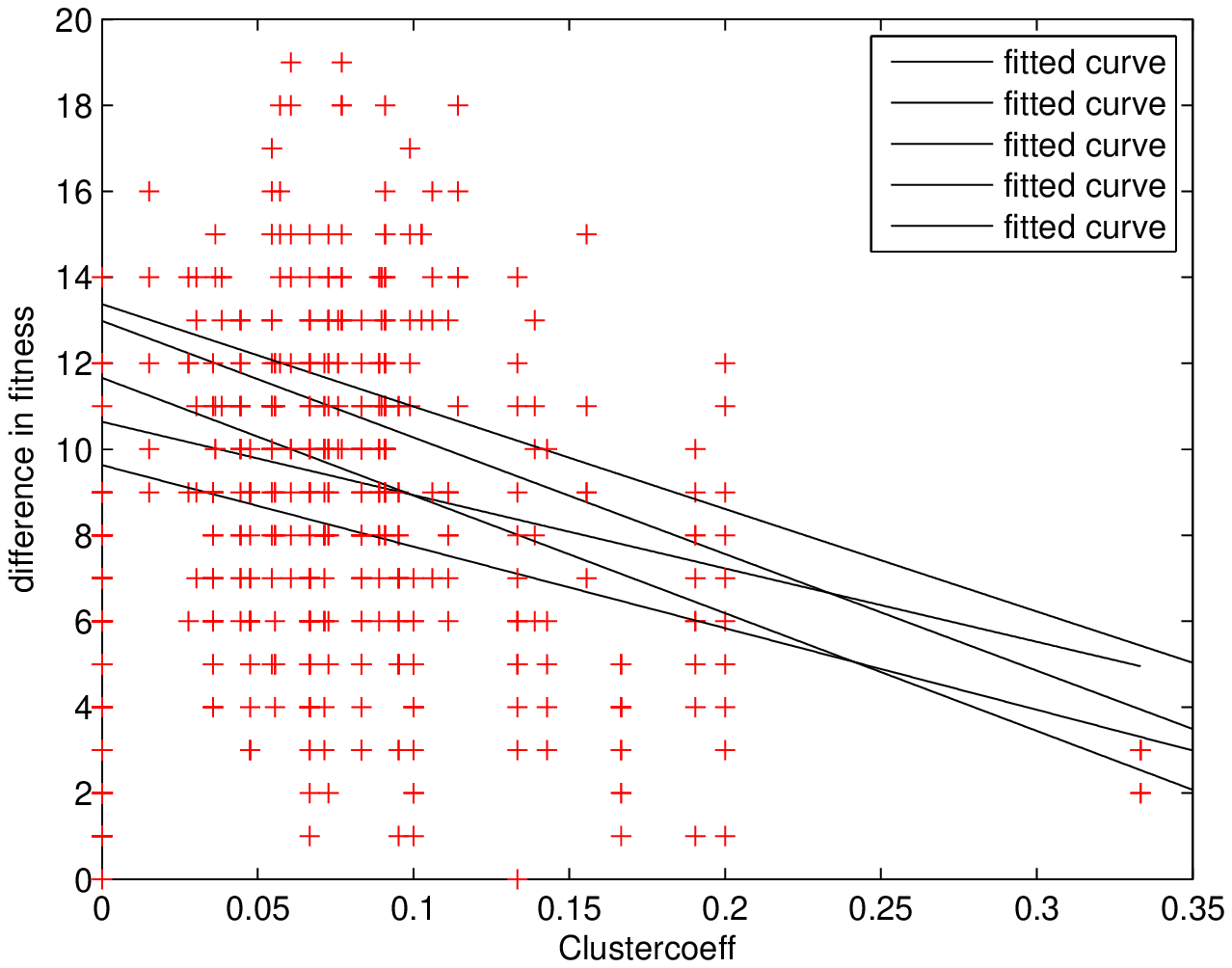}
\caption{Fitness against cluster coefficient in resp hierarchical,non-hierarchical and random network}
\end{center}
\end{figure}

\end{document}